# Energy Level Alignment at Hybridized Organic-Metal Interfaces: The Role of Many-Electron Effects

Yifeng Chen[1], Isaac Tamblyn[2,3] and Su Ying Quek[1,4,*]

[1]Centre for Advanced 2D Materials and Graphene Research Centre, National University of Singapore, 6 Science Drive 2, Singapore 117542
[2]National Research Council of Canada, 100 Sussex Drive, Ottawa, Ontario, K1A 0R6, Canada
[3]Department of Physics, University of Ontario Institute of Technology, Oshawa, Ontario, L1H 7K4
[4]Department of Physics, National University of Singapore, 2 Science Drive 3, Singapore 117551
*: Corresponding author: phyqsy@nus.edu.sg

**ABSTRACT**

Hybridized molecule/metal interfaces are ubiquitous in molecular and organic devices. The energy level alignment (ELA) of frontier molecular levels relative to the metal Fermi level (EF) is critical to the conductance and functionality of these devices. However, a clear understanding of the ELA that includes many-electron self-energy effects is lacking. Here, we investigate the many-electron effects on the ELA using state-of-the-art, benchmark GW calculations on prototypical chemisorbed molecules on Au(111), in eleven different geometries. The GW ELA is in good agreement with photoemission for monolayers of benzene diamine on Au(111). We find that in addition to static image charge screening, the frontier levels in most of these geometries are renormalized by additional screening from substrate-mediated intermolecular Coulomb interactions. For weakly chemisorbed systems, such as amines and pyridines on Au, this additional level renormalization (~1.5 eV) comes solely from static screened exchange energy, allowing us to suggest computationally more tractable schemes to predict the ELA at such interfaces. However, for more strongly chemisorbed thiolate layers, dynamical effects are present. Our ab initio results constitute an important step toward the understanding and manipulation of functional molecular/organic systems for both fundamental studies and applications.

## I. INTRODUCTION

Molecules in both monolayer and single molecular forms constitute basic building blocks for molecular and organic electronics.[1-4] The energy level alignment (ELA) between the metal electrode Fermi level ($E_F$) and frontier molecular orbital (MO) levels determines the energy barrier faced by electrons tunneling across the interface, thus having a critical impact on electronic and charge transport properties.[5-6] For example, a change in the ELA of about 1 eV can change the conductance in single-molecule junctions by an order of magnitude.[7-8] Experimentally, the ELA at molecule/metal interfaces can be determined by photoemission experiments, but correlating the ELA with atomistic-scale knowledge of the interface geometry can be challenging. To directly relate atomic geometries with ELA is highly desirable, underscoring the need to develop accurate theoretical methods to predict the ELA at such interfaces, as well as to develop a basic understanding of the essential physics governing the ELA. However, quantitative prediction of the ELA is nontrivial for hybridized molecule/metal interfaces where wavefunctions are spatially distributed across the interface. Molecules and metals have very different electronic structures, and it is difficult to treat both on the same footing while retaining their respective accuracy. Density functional theory (DFT) methods are routinely unable to predict the correct ELA, and also give incorrect wavefunctions due to inaccurate mixing of the molecule and metal states. Local and semi-local approximations to the exchange-correlation functional typically underestimate molecular HOMO-LUMO gaps by several eVs. Hybrid functional calculations reduce this error, but the optimal percentage of exact exchange in the functional for molecules is different from that for metals. Many-electron GW calculations can correct the band gap problem in DFT, predicting quantitatively accurate band gaps for semiconductors [9-10] and HOMO-LUMO gaps for molecules.[11] Unlike many hybrid functionals, the GW calculations do not rely on empirical fits, and also include the self-energy effects due to screening from the environment.[12-14] In contrast to DFT (a mean-field approach), many-electron GW calculations also provide ab initio insights into the electron self-energy effects at the interface. However, GW calculations are computationally prohibitive for large systems. As a result, very little has been done to understand from first principles the many-electron self-energy effects governing the ELA at hybridized molecule/metal interfaces.

Electron transport in molecule/metal interface systems has been intensively investigated experimentally, both for molecular self-assembled monolayers (SAMs) and single molecular systems. These studies are performed in the context of SAM junctions, realized by contacting the monolayer with a top electrode,[15-16] and single-molecule junctions in scanning tunneling microscopy based experiments or mechanically-controlled break junction experiments,[7-8, 17-20] all of which involve hybridized molecule/metal interfaces, where the MOs of interest overlap with metal states. DFT calculations at the level of semi-local exchange-correlation functionals have provided much insight into these systems, showing the ELA to be affected by a complex interplay between charge transfer, local charge rearrangements, bond dipoles, depolarization fields and Pauli pushback effects.[21-26] On the other hand, it has been shown that DFT with standard exchange-correlation functionals cannot account for all the physics at molecule/metal interfaces. Specifically, for physisorbed molecule/metal interfaces (which do not have hybridization effects), many-electron GW calculations have identified a large substrate-induced renormalization of the MO levels, that has been attributed to non-local static image charge screening from the substrate.[13-14, 27] This information has enabled the prediction of ELA at physisorbed molecule/metal interfaces using a simple two-step DFT+Σ approach, which combines gas-phase self-energies with substrate-induced image charge energies.[13-14, 27-28]

Despite the success of the simple picture of image charge screening known for physisorbed systems, hybridization can potentially change the many-electron screening effects significantly. At hybridized molecule/metal interfaces, molecules form chemical bonds with the metal substrates, with possibility of charge rearrangements. The wavefunctions of the combined system become a superposition of molecular orbitals and metal states. If bonds in the molecule are broken upon chemisorption, this stronger hybridization complicates the picture further. What are the many-electron screening effects at hybridized molecule/metal interfaces? How does hybridization change the picture of static image charge screening? Is it possible to predict the ELA at hybridized molecule/metal interfaces without computationally expensive GW calculations?

Here, we shed light on the many-electron exchange and correlation effects (*i.e.* electronic self-energy effects) on the ELA at hybridized molecule/metal interfaces by performing state-of-the-art GW calculations for prototypical small π-conjugated molecules on Au(111) with common anchoring groups (amine, pyridine and thiol) ), in eleven different geometries. Comparison is made with the DFT+Σ (image charge screening) approach to facilitate our analysis of the many-electron screening effects at these interfaces. In particular, as a measure of the self-energy effects that go beyond gas-phase exchange/correlation and simple image charge screening, we define Λ as the difference between the ELAs obtained with GW and DFT+Σ (magnitude of GW level minus DFT+Σ level). Interestingly, we find that Λ is very large (close to 1.5 eV) for molecular layers, but Λ is close to zero when intermolecular Coulomb interactions are removed in the evaluation of the self-energy ("single molecule limit"). The large Λ for molecular layers brings the frontier levels much closer to $E_F$, and the resulting net GW self-energy correction for hybridized molecular layers on Au is relatively small (less than 0.5 eV) compared to that in physisorbed molecule/metal layers.[13-14, 27] However, this ~0.5 eV correction can still result in a significant change in the frontier MO level relative to $E_F$, as it can remove apparent Fermi level pinning observed at the DFT level for thiolates on Au. We note that the large Λ also results in a smaller HOMO-LUMO gap for the molecular layers, compared to that predicted from image charge screening. On the other hand, the close to zero Λ for the "single molecule limit" is consistent with the success of the DFT+Σ approach in predicting the conductance of single molecule junctions.[7-8, 29-31] This shows that static image charge effects can account for most of the substrate-induced self-energy effects in single molecule junctions, even in the presence of hybridization. The additional level renormalization (beyond static image charge screening) for molecular layers is thus related to intermolecular Coulomb interactions in the evaluation of the self-energy of the hybridized system. The physical origin of this additional renormalization can be traced to static screened exchange for amines and pyridines on Au. This result suggests a way to predict the ELA at such interfaces without GW calculations. However, the self-energy effects are more complicated for the more strongly bound thiolate/Au interfaces, where bonds in the molecule are broken upon adsorption.

## II. METHODS

### A. Density Functional Theory calculations

To relax our hybridized molecule-metal slab geometries, we have used both PBE-D2 [32] and vdW-DF2 [33-34] exchange correlation functionals, to take into account van der Waals interactions. We have used both functionals to relax Geometry **1** and note that the GW ELA is not sensitive to the choice of PBE-D2 versus vdW-DF2 for geometry optimization. Geometry **3** and systems with thiols are relaxed fully with vdW-DF2, the bipyridine geometry is relaxed with PBE-D2, while up-right amines (Geometries **4**, **5** and **6**) are constructed using the bond length and angle parameters of fully relaxed Geometry **3**. All geometrical parameters as well as all atomic coordinates are summarized in Table S8 and Supplementary Section 12.

After geometry optimization, the mean-field DFT eigenvalues and wavefunctions are all computed using the PBE exchange correlation functional with a 19 electron norm-conserving pseudopotential for Au. These eigenvalues and wavefunctions are used as the starting points for both the GW and DFT+Σ calculations.

### B. GW calculations

The many-electron self-energies for the quasiparticle eigenvalues are calculated within the GW approximation.[9, 35] The GW approximation is based on perturbation theory, where a correction to the quasiparticle value is computed from a mean-field DFT starting point. The GW self-energy can be expressed as a sum of two terms, the screened exchange term and the Coulomb-hole term. Both of these terms can in turn be written as a function of the mean-field DFT eigenvalues and wavefunctions. [36]

In a typical one-shot GW calculation, the DFT wavefunctions and eigenvalues are used directly in the expressions for computing the GW self-energies. However, the DFT wavefunctions can be poor approximations to the quasiparticle wavefunctions for hybridized interfaces, and in principle, the self-energy matrix $\Sigma(E)$ should be diagonalized to obtain a better approximation of the quasiparticle wavefunctions, which would constitute a new starting point for perturbation theory (beyond one-shot GW calculations). The diagonalization of $\Sigma(E)$ is a non-trivial task due to its energy dependence. Thus, there are few reports of GW calculations for hybridized molecule/metal interfaces. Here, instead of diagonalizing Σ(E), we follow ref 37 and evaluate the self-energy in the basis of the molecular orbitals, as follows:

$$E_{mol}^{QP} = E_{mol}^{GGA} + \langle \varphi_{mol} | \Sigma(E) - V_{xc}^{GGA}(E) | \varphi_{mol} \rangle. \quad (1)$$

We note that this is still a first-order perturbation theory approach. The motivation for our methodology is that the self-energy correction for metal states is negligible compared to that for molecular states, so that the self-energy operator Σ(E) is expected to be diagonal in the basis of molecular orbitals. (Here, we neglect the matrix elements in Σ(E) involving cross terms between the molecule and metal states, assuming these to also be small compared to the diagonal terms for the molecular orbitals.)

Within typical GW implementations, evaluating (1) requires the projection of the slab wavefunctions onto the molecular orbitals. Since we are investigating hybridized systems, where the molecular projected density of states (PDOS) plots show broadened molecular orbital peaks, the projection of slab wavefunctions onto the molecular orbitals is also helpful to define a single energy for the molecular orbital. Here, we define $E_{mol}^{GGA}$ to be the energy level corresponding to the wavefunction with maximum projection on the MO. This method yields $E_{mol}^{GGA}$ close to the peak maxima for the MOs in the PDOS plots, as shown in Figure S6.

We validate our procedures by: (1) confirming that the self-energy operator is indeed approximately diagonal in the molecular basis (Tables S3-S5), and (2) showing that the GW self-energies match exactly those computed directly using GW, for states that are isolated completely within the molecular region.

GW self-energies involve long-range Coulomb interactions. Due to the 3D periodic boundary conditions used in the calculation, it is standard to use truncated Coulomb interactions in the evaluation of the self-energy for calculations with molecular or slab geometries.[36] The truncation of the Coulomb interactions affects only the self-energy evaluation and does not change the mean-field properties of the system, as evaluated within DFT. In this work, we use a "slab truncation" for the molecular layers, i.e. the Coulomb interaction is truncated in the direction normal to the slabs, so that the artificial presence of other slabs does not change the computed self-energies. To investigate the effect of intermolecular Coulomb interactions on the self-energy, we have also employed "box truncation", where the Coulomb interaction is truncated in all three spatial directions. We note that this truncation does not affect the intermolecular Coulomb interactions in the DFT mean field calculations (including dipole-dipole interactions), or the vacuum potential; both are unchanged throughout the calculation.

C. DFT+Σ approach (static image charge screening)

The two-step DFT+Σ approach is introduced here to facilitate analysis of the self-energy effects. This approach adds a self-energy correction Σ to the DFT ELA for the molecule/metal system, evaluating Σ as the sum of two components: (1) the self-energy effects in the gas phase isolated molecule, and (2) the self-energy effects caused by the substrate, which is taken to be equal in magnitude to the substrate-induced image charge energy when an electron is added to or removed from the molecule.[7-8, 30] The DFT+Σ approach is based on GW calculations for the physisorbed system,[13] where it was proven numerically and shown analytically, that under certain assumptions, the substrate-induced change in self-energy for a single physisorbed molecule on a metal substrate, is given by the image charge energy. Intuitively, one can understand this result as follows. The HOMO and LUMO levels of a molecule measure the amount of energy required to remove or add an electron to the molecule. In the presence of a metal substrate, the resulting hole/electron in the electron removal/addition process will be stabilized by screening from the metal substrate. In the physisorbed case, the screening effect from the substrate can be approximated with classical electrostatics as the image charge energy. This image charge energy makes it easier to remove and add electrons to the molecule, thus reducing the HOMO-LUMO gap. We note, however, that this simple picture ignores other substrate-induced exchange and correlation effects on the ELA.

The image plane position is taken to be 0.9 Å above the Au surface. Mulliken populations on each atom in the molecule are used for computing the image charge energy. The image charge energy thus computed is almost identical to that assuming a point charge in the middle of the molecule, except for Geometry **11** where the sulfur atom (which contains 53% of the total charge) is located very close to the image plane (Table S10).

Further details of methods are given in the supplementary information.

## III. RESULTS AND DISCUSSION

### A. GW convergence of absolute energy levels

The accurate prediction of ELA in a hybrid system requires absolute (rather than relative) convergence of energy levels in the separate component systems; in particular, it is essential for GW to properly describe the Au(111) work function as well as the ionization potential (IP) and electron affinity (EA) of the molecules. While the DFT PBE Au(111) work function matches well with experiment, achieving the same accuracy in GW is non-trivial.[37] However, for the parameters reported in this work (both “slab truncation” and “box truncation”), the work function for Au(111) obtained after GW self-energy corrections matches well with the PBE work functions (Fig. S1). This is partly attributed to the very large 20 Ry cutoff energy for the dielectric matrix. Such a choice is also sufficient to converge the IP and EA for the gas-phase molecules (Table S2). In particular, we note that the benzene-diamine (BDA) IP agrees well with experimental values and previous GW calculations.[38]

We note here that different approximations used to treat the $\mathbf{q} \rightarrow 0$ limit in the GW self-energy calculations for metallic systems [36] result in different rigid shifts of all quasiparticle levels (Table S1). A careful calibration is therefore required for these ELA calculations. We note that the HOMO-LUMO gap (an easier quantity to converge) is computed to be the same in all these different approximations.

B. ELA for benzene-diamine (BDA) molecular layer on Au(111)

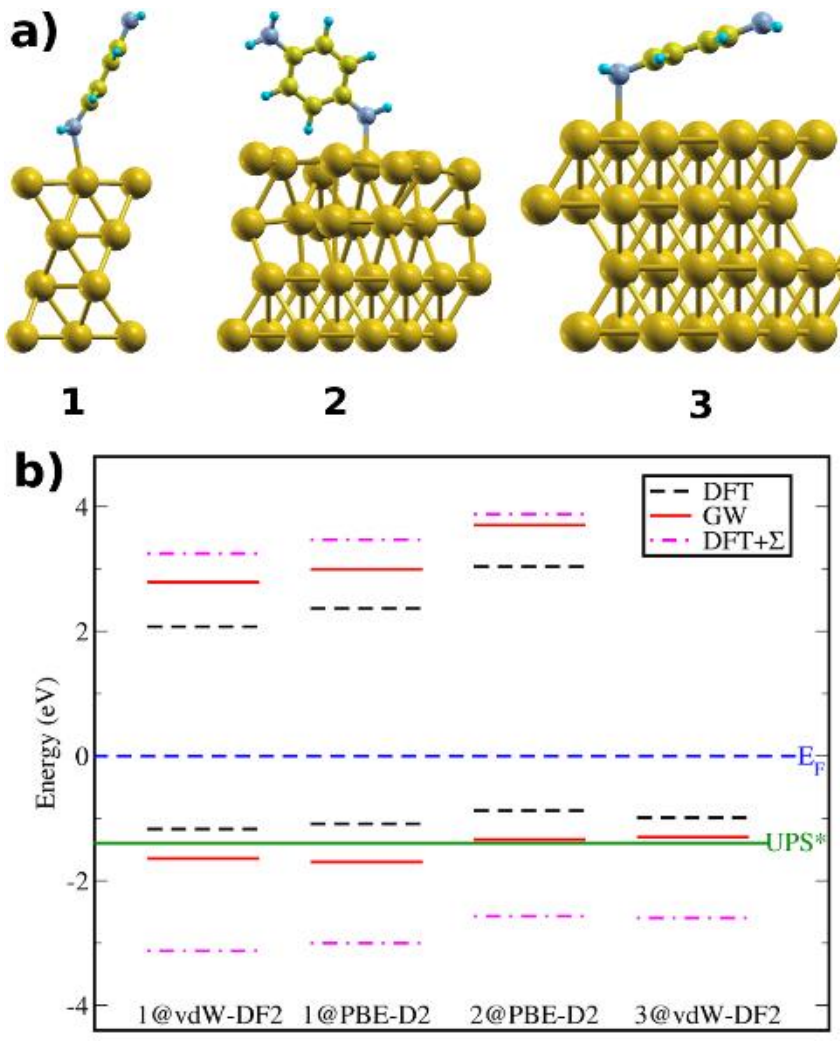

Figure 1: (a) Unit cell for BDA molecular layer on Au(111). Geometry **1**: √3x √3 R30° Au cell. Geometries **2** and **3**: 3x3 cell. The gold, yellow, bright blue and gray balls denote Au, C, H and N atoms, respectively. (b) HOMO/LUMO ELA diagram obtained from GW, DFT and DFT+Σ. The exchange-correlation functional used for geometry optimization is indicated. The experimental UPS HOMO position (-1.4±0.1 eV) is from Reference [39]. Note that the LUMO level for Geometry **3** is not spectroscopically defined (there is no clear peak when slab wavefunctions are projected onto the LUMO).

We begin our discussion with the ELA in BDA-Au(111) systems, where the HOMO alignment is known experimentally.[39] We consider different molecular orientations and coverage, as shown in Figure 1a. Geometries **1** and **2** are tilted structures with different coverage, while Geometry **3**, the face-on configuration, is constructed based on experimental information,[39] and then fully relaxed using the vdW-DF2 functional as implemented in VASP. In the relaxed geometry, the BDA molecule in Geometry **3** is tilted 24° from the Au(111) surface, in excellent agreement with experimental NEXAFS data.[39] Following prescriptions in the literature, the Coulomb interaction is truncated in the direction normal to the slabs during the evaluation of the GW self-energies (later referred as “slab GW”).[36, 40]

We focus first on the GW results, shown in red in Figure 1b. The GW HOMO alignment for Geometry **3** is in good agreement with the experimental ultraviolet photoelectron spectroscopy (UPS) value.[39] The HOMO alignments in Geometries **1** and **2** are reasonably close to the UPS value as well, with very similar values for **1**@vdW-DF2 and **1**@PBE-D2. Moving to the DFT and DFT+Σ ELA (Figure 1b), we see that DFT (with PBE exchange-correlation functional) gives HOMO levels that are ~0.5 eV too shallow compared to GW and UPS results, as expected from the underestimation of HOMO-LUMO gaps with DFT PBE. In contrast, the DFT+Σ HOMO levels are ~1.5 eV deeper than the GW HOMO levels.

We note that our results appear to disagree with previous literature,[39, 41] where the DFT+Σ ELA matched reasonably with the UPS result. There are several differences worth noting here. Firstly, the atomic geometries used above are different from those in References 39 and 41. The DFT ELA is very sensitive to the details of the atomic geometry, thus giving different DFT starting points for the calculation. Using the atomic geometries from the literature, we obtain the same DFT (PBE) HOMO alignment as in the references (-0.4 eV for the BDA molecule, tilted 24° from the Au(111) surface, in a 4x4 cell,[39] and -0.6 eV for the linear chain motifs of BDA on Au(111) [41]). In comparison, our calculations show that Geometry **3**, which also has BDA tilted 24° from the Au(111) surface, but in a 3x3 cell, has a DFT HOMO level of -1.0 eV. The deeper DFT HOMO level for higher coverage is consistent with a net larger electrostatic potential shift for denser coverage due to a higher density of Au-N bond dipoles (upwards with N losing charge to Au).[25] The sensitivity of the DFT HOMO alignment to atomic geometries leads to the question of whether the GW HOMO alignments are equally sensitive, and whether or not the atomic geometries used in this work indeed correspond to the experimental geometry. Our calculations suggest that the GW ELA is fairly robust across Geometries **1**, **2** and **3**, while Reference 37 also gives a similar GW ELA for a BDA molecule physisorbed flat on Au(111), provided that the Fermi level is shifted to account for the error in the Au work function (there, the DFT PBE HOMO level was much shallower at -0.25 eV). On the other hand, we have also performed a GW calculation on the linear chain motif geometry,[41] and found a GW HOMO level of -0.75 eV, which is much closer to $E_F$ than the UPS value. Regarding the choice of geometries, we note that the linear chain motif geometry is stabilized by hydrogen bonds at 5 K,[42] and is thus not likely to be the geometry in the room temperature UPS experiment.[39] Relaxing the BDA molecule in a 4x4 cell results in a tilt angle of 18° (Figure S3) instead of 24° (the experimentally determined tilt angle), suggesting that the 3x3 cell may be more appropriate. (See Section 8 of the SI for more details.)

Another important difference is that here, the image plane position is taken to be 0.9 Å above the Au(111) surface, following a recent report that takes into account quantum mechanical exchange and correlation effects in the determination of the image plane position.[43] In contrast, the previous works used a classical image plane position of 1.47 Å above the Au(111) surface,[39, 41] thus giving much larger image charge corrections. In Reference [27], we found that the quantum mechanical definition of the image plane is required to obtain ELA in good agreement with experimental scanning tunneling spectroscopy data for PTCDA on Au(111).

A comment about surface Au adatoms is in order here. We have also checked tilted and face-on configurations of BDA molecules anchored on Au adatoms on Au(111), and found scattered GW HOMO positions not matching the UPS value, indicating that the BDA molecules are not adsorbed on Au adatoms in the UPS experiment, consistent with the experimental report.[39]

C. Molecular layer versus "single molecules"

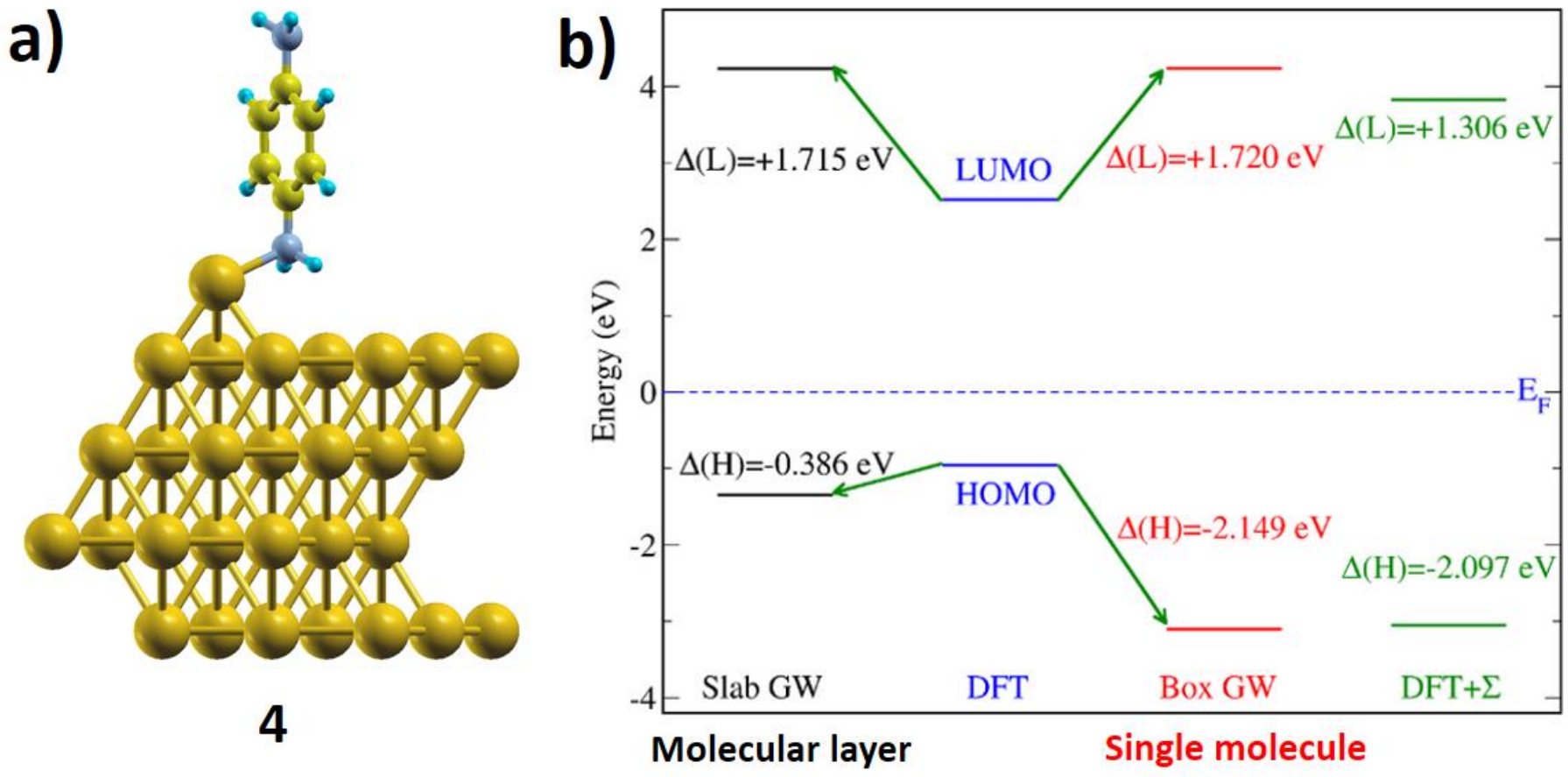

Figure 2: (a) Geometry **4**: BDA oriented upright, bonded to a Au adatom in a 3x3 Au(111) cell. (b) HOMO/LUMO ELA diagram obtained from GW, DFT and DFT+Σ. Slab GW applies for the molecular layer, while box GW models the "single molecule limit".

Since DFT+Σ ELA was able to give a reasonable conductance for single molecule BDA-Au junctions, we study a BDA-Au interface that is more representative of the single molecule junctions, by anchoring BDA on Au adatoms (Geometry **4**; Figure 2a).[44] The GW HOMO level is still 1.7 eV closer to $E_F$ than the DFT+Σ HOMO level. This result is in stark contrast to the excellent performance of DFT+Σ in predicting the conductance values of single-molecule BDA-Au junctions,[7-8] which in turn strongly suggests that DFT+Σ gives a reasonable ELA there (the conductance being extremely sensitive to the level alignment of frontier molecular orbitals relative to $E_F$). Since our GW calculations are performed on geometries at fairly high coverage, we consider what happens if the Coulomb interactions between molecules is truncated when evaluating the GW self-energy (using a Wigner-Seitz box truncation scheme [36, 40]). In that case, we find that the predicted HOMO level deepens by 1.8 eV, and is much closer to the DFT+Σ result (Figure 2b). We call this "box GW" and note that similar truncation schemes have previously been used to describe other extended systems.[45-46] Due to the relevance of these results to single molecule junctions, we also refer to these calculations as modeling a "single molecule". However, we note that our model does not strictly represent a single molecule on the Au surface, because intermolecular effects are already included at the DFT level.

Using box GW for Geometry **2** with no Au adatom, we also find that the box GW and DFT+Σ HOMO levels match reasonably well (Table S11), thus indicating that the agreement between box GW and DFT+Σ does not depend on the presence of the Au adatom.

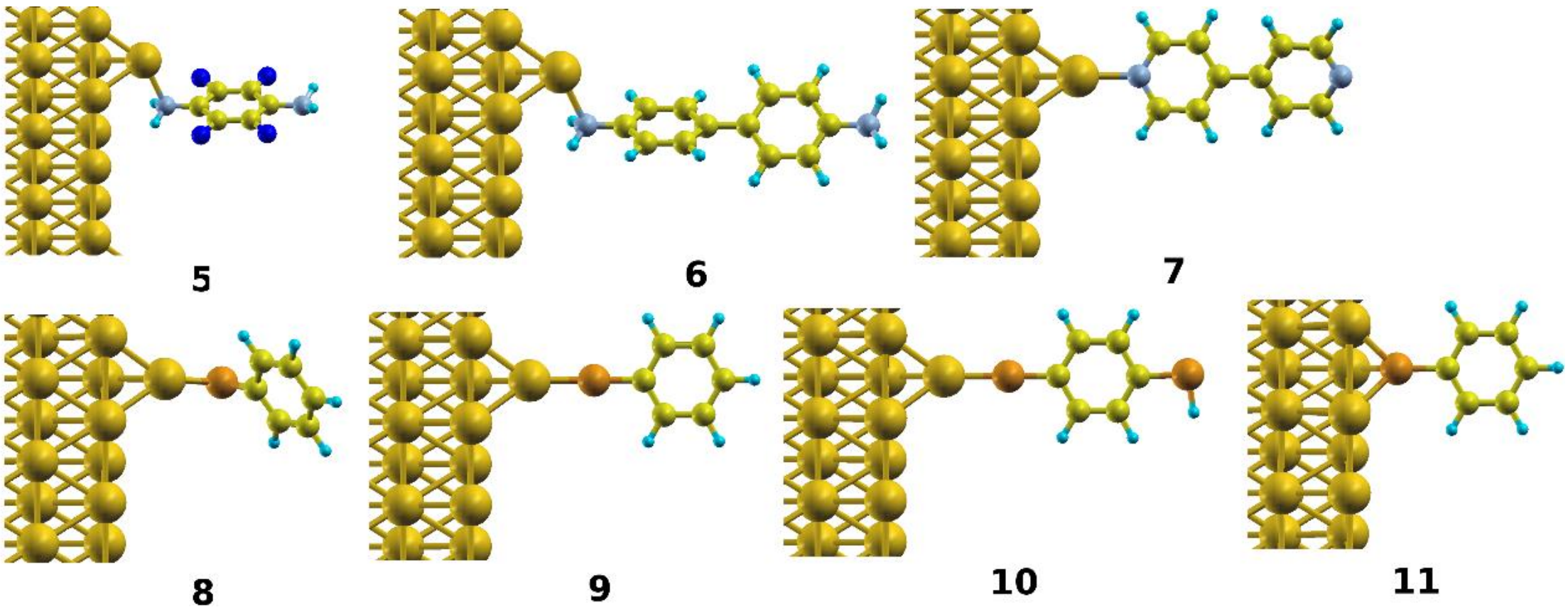

Figure 3: Geometries **5-11.** Different molecules in a 3x3 Au(111) cell. **5**: fluorinated BDA (FBDA), **6**: biphenyldiamine (BPDA), **7**: 4,4'-Bipyridine (BP), **8**, **9**, **11**: benzene-thiol (BT), **10**: benzene-dithiol (BDT). In **5-10**, the molecules are anchored on Au adatoms. In geometry **11**, BT is anchored on fcc hollow site of Au(111). Dark blue balls denote F and orange balls denote S.

To assess the generality of our results and the dependence on the anchoring group, we consider other prototypical molecules, as shown in Figure 3. We focus on the MO levels relevant for transport in the corresponding single molecule junctions (HOMO for amines and thiols, LUMO for pyridines). These anchoring groups have all been used extensively in single-molecule junctions, and thiols are also the anchoring group of choice in SAMs.[47] While the binding geometries for amines and pyridines are well-known, those for the more strongly interacting thiol-Au systems are surprisingly controversial.[48-49] Our choice of motifs for thiols on Au are far from exhaustive. The general consensus is that the S-H bond in the molecule cleaves upon adsorption on Au, forming a thiolate. This creates difficulties in our projection procedure; we find that using the radical orbitals as projectors leads to unphysical results, and instead use the original thiol molecule for obtaining the MO projectors, which gives reasonable results (see Figure S6 for comparison of the projection method with the molecular PDOS). Bader charge analysis reveals a substantial intra-molecular charge re-organization upon hydrogen cleaving and molecular adsorption, with the sulfur atom gaining an extra ~$1.4e^-$ charge (Table S7). Despite the charge rearrangement, the HOMO levels of the thiolate molecules are not mixed with other MOs in the geometries considered here.

The level alignment results are shown in Table 1 and Figure 4. Remarkably, we find that for all these systems, slab GW gives a level alignment that is significantly closer to $E_F$ than DFT+Σ, indicating a large renormalization Λ coming from screening effects beyond simple image charge screening. In comparison with slab GW, Λ is very small in box GW, except for Geometry **11** (benzenethiol anchored on the fcc hollow site) where Λ is 1.5 eV. The stark contrast between Λ's computed for slab GW and box GW indicates that the additional screening effects in the molecular layer are related to intermolecular screening. Besides level alignment, we have also compared the molecular HOMO-LUMO gaps obtained with slab GW, box GW and DFT+Σ. Except for Geometry **11**, the box GW and DFT+Σ gaps agree reasonably well, while the slab GW gaps are ~1 eV too small.

Table 1: MO levels, referenced to $E_F$, for Geometries **5-11** (units in eV).

| | FBDA HOMO @Geo.**5** | BPDA HOMO @Geo.**6** | BP LUMO @Geo.**7** | BT HOMO @Geo.**8** | BT HOMO @Geo.**9** | BDT HOMO @Geo.**10** | BT HOMO @Geo.**11** |
|---|---|---|---|---|---|---|---|
| DFT | -1.18 | -0.76 | +0.18 | -0.27 | -0.04 | 0.00 | -1.26 |
| DFT+Σ | -3.33 | -2.70 | +1.51 | -2.56 | -2.41 | -2.06 | -2.07 |
| Box GW | -3.51 | -2.94 | +1.68 | -2.22 | -2.48 | -2.55 | -3.60 |
| Slab GW | -1.71 | -1.17 | +0.40 | -0.72 | -0.27 | -0.34 | -1.80 |

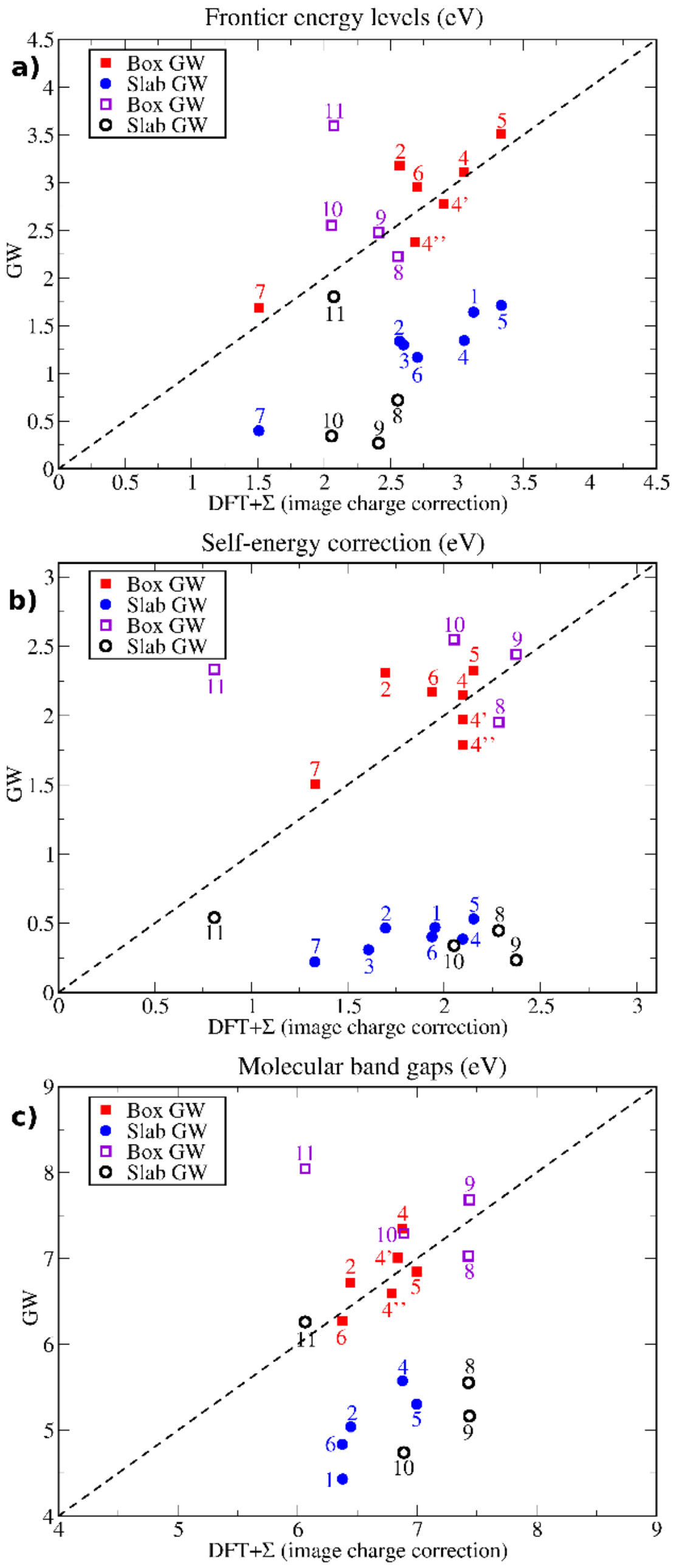

Figure 4: Scatter plots of (a) absolute frontier orbital energy levels relative to $E_F$, (b) magnitude of self-energy correction relative to DFT levels, (c) molecular HOMO-LUMO gaps: slab/box GW versus DFT+Σ, indexed by geometry. The empty symbols in different colors denote thiolate/Au systems.

We also investigate two-sided metal/molecule/metal junction geometries for BDA with two different coverages, which are more representative of single molecule junctions (Figure S4). For the √3x √3 R30° cell, we found a shallower slab GW HOMO level at -1.48 eV, compared to a corresponding HOMO of -1.78 eV for the one-sided interface. Both are much shallower than the DFT+Σ HOMOs of -2.95 eV (two-sided) and -3.24 eV (one-sided). The box GW and DFT+Σ HOMO levels are also quite close in the 3x3 cell junction (-2.15 eV and -2.32 eV, respectively).

Thus, for all geometries considered here, it is possible to draw two general conclusions. First, it is clear that there is a large additional energy renormalization for molecular orbital levels in chemisorbed molecular layers on Au(111), that go beyond the picture of static image charge screening. Second, the box GW and DFT+Σ levels are in reasonable agreement, although with poorer agreement for the thiol in Geometry **11**. This suggests that static image charge screening can account for a large part of the substrate-induced electron self-energy effects in "single molecule" junctions, explaining why DFT+Σ models could predict the electronic conductance in these systems.[7-8, 29-31]

D. Many-electron effects at hybridized molecule/metal interfaces

*1. "Single molecule limit"*

Given the above conclusions, a natural question to ask is why box GW and DFT+Σ levels agree reasonably well, and whether or not this is a general result. To answer this question, it is instructive to revisit the following assumptions used in the analytical derivation of the DFT+Σ formalism for physisorbed single molecules on metal substrates:[13] (I) The molecular orbitals do not overlap with the metallic states; (II) charge transfer between the molecules and the metal should be negligible; (III) molecular polarizability is neglected; (IV) substrate-induced dynamical self-energy effects are neglected; (V) the change in the screening potential upon molecular adsorption can be approximated by an image charge form. To what extent do the hybridized systems satisfy these assumptions?

We can show that Assumption (I) may be relaxed to include hybridized molecule/metal systems in which the molecular orbitals (MOs) of interest do not mix with other MOs upon hybridization with the metallic states (slab wavefunctions with non-zero projections on the MO of interest should not have non-zero projections on other MOs). This is an important criterion and is found to be satisfied for all the MOs considered in Figure 4. The charge transfer between the molecule and the metal is negligible in our systems (Table S7). The effects of molecular polarizability are also small. The small molecular polarizability can be inferred from GW calculations for the molecular layer without the metal substrate, where the GW gap (8.0 eV) is very close to that of the single molecule (8.2 eV). We also note that even when the intra-molecular polarizability is large (such as for a PTCDA monolayer on Au(111)), the effect of molecular polarizability on the ELA is only a few tenths of eV.[27]

Assumptions (IV) and (V) are more difficult to verify without GW calculations on the full system. In particular, one should not expect DFT+Σ to give a high degree of quantitative accuracy in hybridized systems. However, it is clear from Figure 4b that a significant improvement over regular DFT calculations can be achieved, with self-energy corrections between 1.5-2.5 eV. Geometry 11 (thiol on hollow site) is an exception; we can trace the disagreement to an overly large image charge energy that results from the sulfur atom being very close to the Au(111) surface (charge rearrangements result in the sulfur atom carrying significant charge; Table S7). We have also performed additional box GW calculations for Geometry **4**, with another two layers of Au (**4'**), and also with a larger 4x4 Au cell (**4"**) (slab GW calculations are too expensive for these geometries). The results are shown in Figure 4, with reasonable agreement between GW and DFT+Σ. Thus, static image charge effects can account for a large part of the substrate-induced self-energy for many hybridized single molecule/metal systems (Figure 4), providing a computationally efficient method (DFT+Σ) to approximate the ELA in such systems.

*2. Substrate-induced intermolecular self-energy effect*

What is the origin of the large additional level renormalization, beyond image charge screening, for molecular layers on Au(111)? Since this additional renormalization is absent in box GW calculations, we expect that intermolecular Coulombic interactions play a key role. We first ask if the intermolecular interactions could have resulted in additional band dispersion in GW. From Table S9, we can see that the HOMO energies at four high symmetry k-points are essentially the same, suggesting that both DFT and GW HOMO levels are flat. Thus, the additional level renormalization affects the molecular orbital levels at all k-points equally, and is likely to come from long-range intermolecular screening. On the other hand, the almost identical GW gaps for the isolated molecule and the molecular layer point to the importance of the metal substrate in mediating the additional screening.

To shed more light on the origin of this effect, we analyze the GW self-energies in terms of its two distinct components (screened exchange and Coulomb-hole terms) [9, 36] (Table 2). The screened exchange term is the exchange interaction with the bare Coulomb interaction replaced by the screened Coulomb interaction, while the Coulomb-hole term is the interaction of the quasiparticle with the induced potential due to the rearrangement of electrons around the quasiparticle.[9] We also analyze the relative importance of static *versus* dynamical terms in the substrate-induced self-energy corrections by performing static COHSEX calculations [36] (which computes the screened exchange and Coulomb-hole self-energies in the static limits) and comparing those with the GW results.

*(a) Hybridized systems with "weak" chemisorption*

The data is qualitatively different for amines/pyridines compared to the more strongly bound thiols. Here, we associate the amines/pyridines with "weak" chemisorption, where no bonds are broken and the charge rearrangement upon adsorption is negligible. In contrast, the S-H bond in thiols are broken, leading to large intramolecular charge rearrangements - "strong" chemisorption. This distinction between "weak" and "strong" chemisorption scenarios has also been discussed in previous literature.[46, 50] We focus first on the former, where the Coulomb hole term is essentially unchanged between the slab GW and box GW results (green; Table 2), so that the change in the GW self-energy, going from the "single molecule limit" to the molecular layer, comes solely from the screened exchange term, which moves the frontier levels closer to $E_F$ for the molecular layer (red versus blue; Table 2). In contrast, the screened exchange terms (and the Coulomb hole terms) in the isolated BDA layer are similar for slab GW and box GW calculations (Table 2), further confirming that it is the presence of the Au substrate that renormalizes the intermolecular screening. We also see that most of the dynamical effects (differences between GW and COHSEX) come from the Coulomb-hole term, and that the screened exchange term is essentially the same in the GW and COHSEX calculations (red and blue; Table 2). Thus, the additional screening observed for "weakly" chemisorbed molecular layers is a static effect.

These observations allow us to propose a scheme (Figure 5a) to estimate the ELA in hybridized molecule/metal systems with "weak" chemisorption, such as when no bonds are broken, and no large charge rearrangements occur. An example would be chemisorption involving only donor-acceptor interactions (such as in amines/pyridines on Au). First, the ELA in the "single molecule limit" ("box GW") can be estimated using the DFT+Σ static image charge approach, as suggested by Figure 4. Next, one can account for the additional level renormalization for the molecular layer ("slab GW" *versus* "box GW") by taking the difference between the static screened exchange terms computed using slab and box Coulomb truncation schemes (difference between red and blue in Table 2). While being just an estimate compared to the fully *ab initio* GW calculations, the proposed scheme is physically motivated by the data presented in this work, and is computationally less expensive than GW, since the static screened exchange term involves only occupied states and is much easier to compute than the dynamical GW self-energy.[51] The results are shown in Figure 5b.

Table 2: Components of the GW and static COHSEX self-energy corrections for different geometries, together with the bare exchange term (X). SX is screened exchange, CH the Coulomb hole term. The GW self-energy is equal to the difference between (SX+CH) and the mean-field exchange-correlation term. The X term is given for comparison. Units are in eV, given to 1 d.p. for clarity. (More data in Table S6)

| | Method | X | SX | CH |
|---|---|---|---|---|
| HOMO for isolated BDA monolayer (no Au), with structure as in Geo. **4** | Slab GW | -17.5 | -10.2 | -7.7 |
| | Box GW | -17.6 | -10.5 | -7.1 |
| BDA HOMO in | Slab GW | -15.9 | -7.3 | -8.0 |
| | Slab | -15.9 | -7.2 | -9.4 |
| Geo. **4** | static-COHSEX | | | |
| | Box GW | -17.5 | -9.2 | -7.9 |
| | Box static-COHSEX | -17.5 | -9.1 | -9.7 |
| BP LUMO in Geo. **7** | Slab GW | -13.8 | -5.4 | -8.4 |
| | Slab static-COHSEX | -13.8 | -5.2 | -9.5 |
| | Box GW | -9.7 | -4.0 | -8.5 |
| | Box static-COHSEX | -9.7 | -4.2 | -9.4 |
| BT HOMO in Geo. **9** | Slab GW | -13.6 | -4.5 | -8.5 |
| | Slab static-COHSEX | -13.6 | -4.1 | -9.9 |
| | Box GW | -15.4 | -7.6 | -7.6 |
| | Box static-COHSEX | -15.4 | -7.5 | -9.3 |

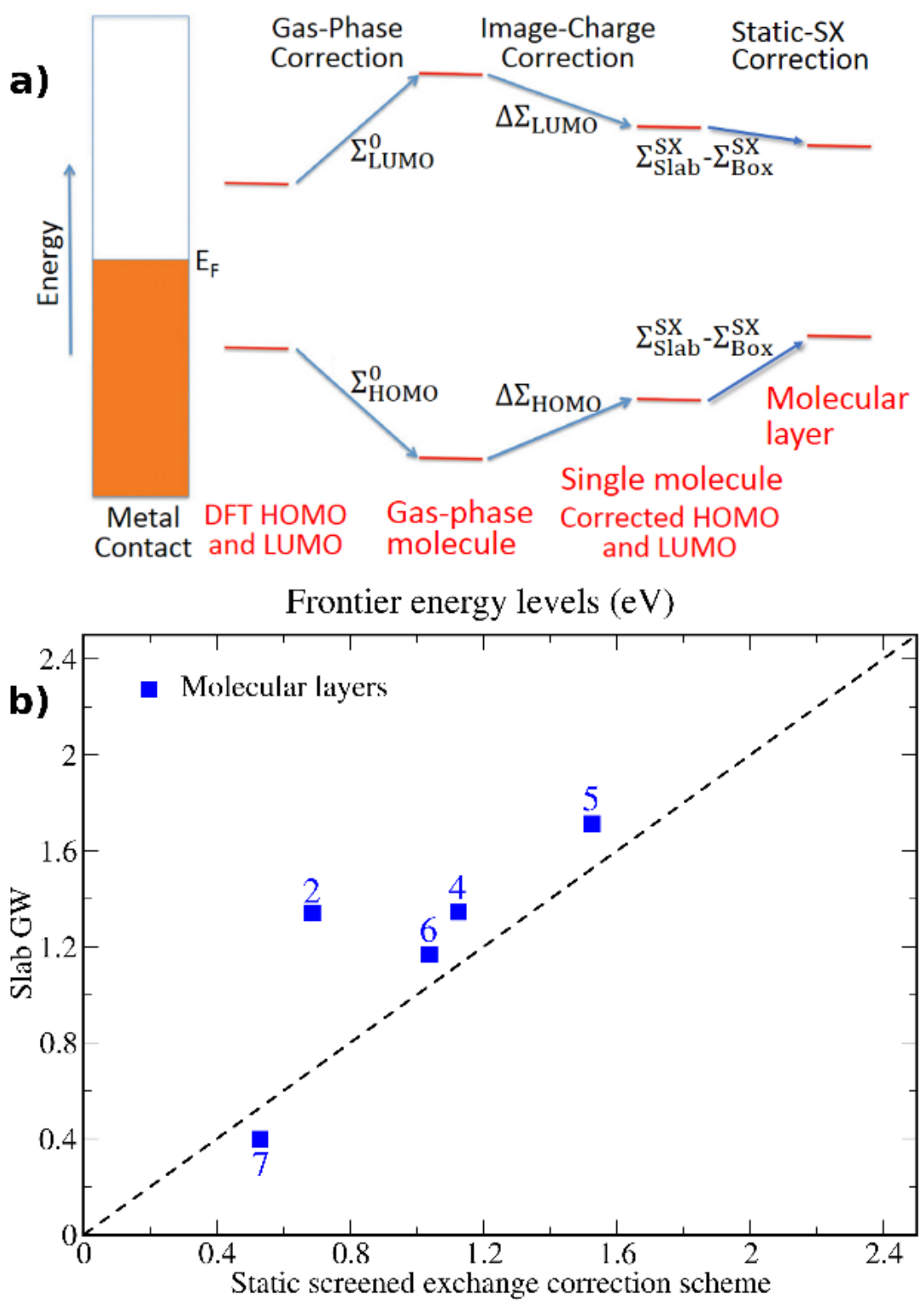

Figure 5: (a) Static screened exchange (SX) correction scheme upon the two step DFT+Σ method for molecular layer ELA on metal substrates. (b) Slab GW levels *versus* static SX corrected molecular layer energy levels. The screened exchange correction scheme only applies in the "weak" chemisorption case for amines and pyridine.

*(b) Hybridized systems with "strong" chemisorption*

For thiols, where the S-H bond in the molecule is broken upon adsorption on Au, the picture is more complex. The screened exchange term also moves the HOMO closer to $E_F$ in slab GW compared to box GW; however, this is partially compensated by the Coulomb-hole term moving the HOMO in the opposite direction. This clearly shows that the self-energy effects for the more strongly bound thiolate/Au systems are qualitatively different from those of amines and pyridines on Au. Dynamical effects are also found to be significant.

## IV. CONCLUSIONS

Using state-of-the-art GW calculations with a projection approach, we have shown that the electron self-energy effects at hybridized molecule/metal interfaces are non-trivial. Our GW results on more than ten geometries provide *ab initio* data to understand the physical origin of many-electron effects on the ELA at hybridized molecule/metal interfaces with "weak" and "strong" chemisorption. This physical understanding allows us to suggest computationally less expensive schemes to estimate the ELA for hybridized molecule/metal interfaces, except for the case of strongly chemisorbed molecular layers. Thus, despite an expected tradeoff in quantitative accuracy, the proposed schemes can be used to obtain reasonable estimates for a large class of hybridized molecule/metal interfaces more efficiently, paving the way toward high throughput analysis of structure-ELA relationships in such interfaces. The new insights into many-electron screening effects at hybridized molecule/metal interfaces also constitute an important step towards the understanding and manipulation of functional molecular/organic systems for both fundamental studies and applications.

We find that when intermolecular Coulomb interactions are absent in the evaluation of the GW self-energy, the GW ELA agrees reasonably well with that predicted by the DFT+Σ image charge approach, for amines and pyridines on Au, and for some thiolate-Au geometries. This remarkable agreement explains the excellent performance of the DFT+Σ approach in predicting the conductance of single-molecule junctions.[7-8, 29-31] On the other hand, when intermolecular interactions are included, we uncover a huge (close to 1.5 eV) substrate-induced intermolecular screening effect that brings the frontier levels much closer to $E_F$. Good agreement is achieved with experimental UPS data for a BDA layer on Au(111). For "weak" chemisorption (amines and pyridines on Au), this substrate-induced intermolecular screening effect is a static effect manifesting itself in the screened exchange-interaction term. For "strong" chemisorption (thiolate/Au systems), substrate-induced dynamical self-energy effects and changes in the Coulomb-hole term are also important. As a final remark, we note that our results imply that if one were to compute the conductance of single molecule junctions using GW, one has to be careful to truncate the Coulomb interactions between periodic copies of the molecules. In the literature,[52-54] the GW calculation is performed only for a finite region of the junction without including intermolecular screening effects. This also explains why the GW conductance was in reasonable agreement with experiment.

## Supporting Information

The Supporting Information is available free of charge on the URL.

Details of computation methods, GW projection, GW convergence of metal and molecule, GW self-energy off-diagonal elements, other geometries, decomposed GW versus static COHSEX, Bader charge analysis, binding energy, geometric parameters (including full atomic coordinates), DFT and GW bandstructure for isolated BDA layer, GW MO levels across the Brillouin zone, image charge correction amount, molecular PDOS, and table of all data.

## ACKNOWLEDGMENTS

S.Y.Q. and Y.C. acknowledge support from the Singapore National Research Foundation, Prime Minister's Office, under its medium-sized centre program and under grant NRF-NRFF2013-07. I.T. acknowledges support from NSERC and SOSCIP. Computations were performed on the NUS Graphene Research Centre cluster. We thank Guo Li and Jeff Neaton for sharing the BDA linear chain model reported in the supporting information. We thank Keian Noori for discussions on related work.

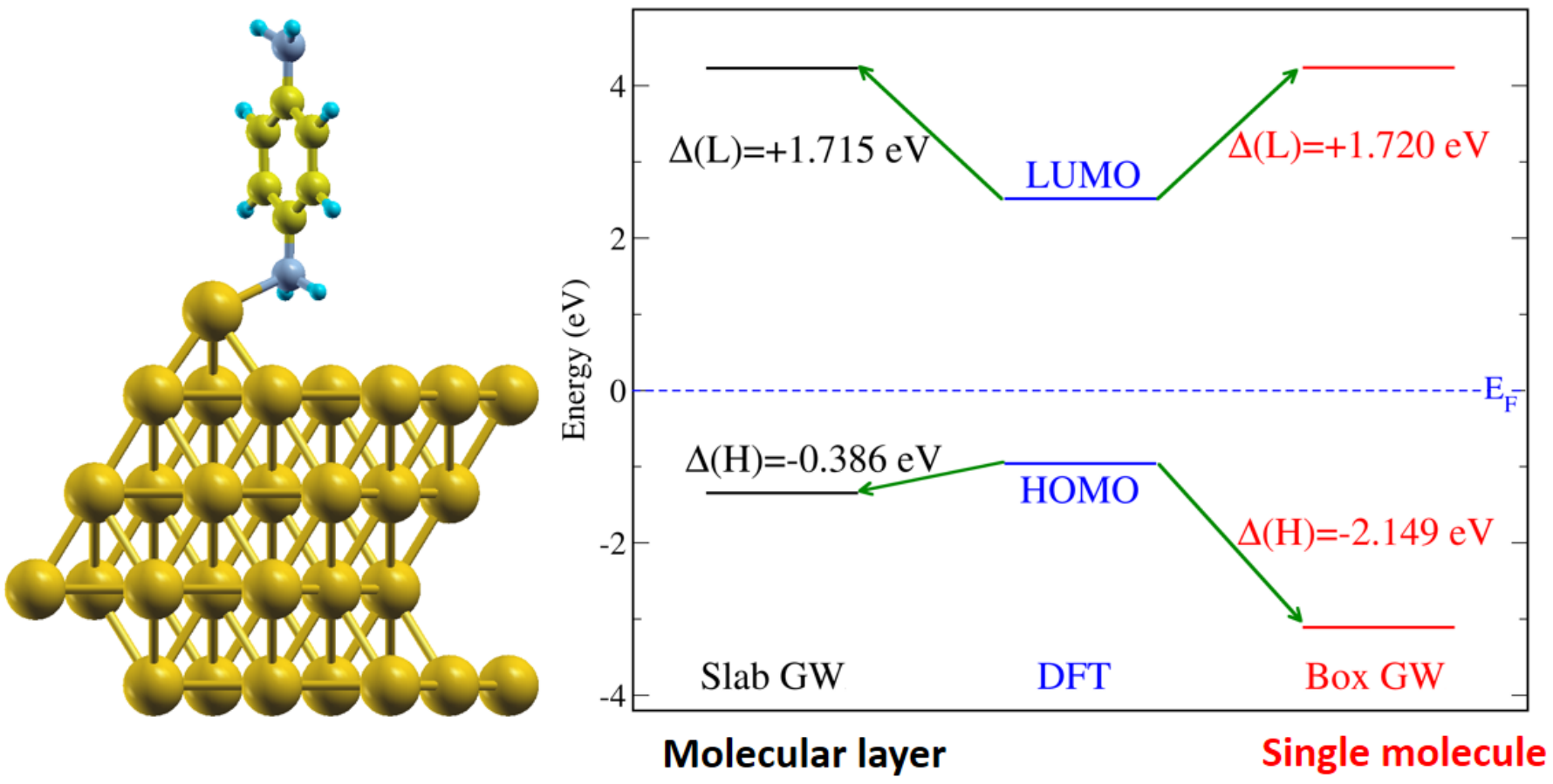

# Table of Contents Image

## Supporting Information for

# Energy Level Alignment at Hybridized Organic-Metal Interfaces: The Role of Many-Electron Effects

Yifeng Chen[1], Isaac Tamblyn[2,3] and Su Ying Quek[1,4*]

[1]Centre for Advanced 2D Materials and Graphene Research Centre, National University of Singapore, 6 Science Drive 2, Singapore 117542

[2]National Research Council of Canada, 100 Sussex Drive, Ottawa, Ontario, K1A 0R6, Canada

[3]Department of Physics, University of Ontario Institute of Technology, Oshawa, Ontario, L1H 7K4

[4]Department of Physics, National University of Singapore, 2 Science Drive 3, Singapore 117551

*: Corresponding author: phyqsy@nus.edu.sg

**Table of contents**

**12) Molecule/metal slab geometric parameters**

**13) DFT and GW bandstructure for the BDA layer as in Geo. 4**

**14) GW projected molecular level dispersion across the Brillouin zone**

**15) Calculated image charge correction energy for the hybridized slabs**

**16) Projected density of states together with our DFT starting points marked for Geo.'s 1-11 investigated in this work.**

**17) Table S11 summarizing all data**

## 1) General details on computational methods used

Geometry optimization was performed using both VASP [1] and Quantum-ESPRESSO [2] (Q.E.). To account for long-range van der Waals interactions, we used both the Grimme's PBE-D2 method [3] (as implemented in Quantum Espresso) and the van der Waals density functional (vdW-DF2, optB86b) [4,5] (as implemented in VASP); key geometric parameters are given in Table S8 for all geometries in this work. Geometry **3** was built from experimental information [6] and relaxed in VASP with vdW-DF2 (optB86b), while the BDA linear chain model was provided by Guo Li [7]. Bader charge analysis was performed on total charge densities produced by Q.E. with the UTexas code [8].

GW calculations were done using the BerkeleyGW package [9], with DFT wavefunction input taken from Q.E. using a norm-conserving pseudopotential, PBE exchange-correlation functional, 60 Ry kinetic energy cutoff, and $10^{-6}$ Ry convergence threshold. Due to the spatial overlap of semi-core and valence wavefunctions in Au [10], we used a 19-electron Au pseudopotential in our GW calculations. A vacuum size of at least 13 Å between periodic slabs was used, and the Coulomb interactions between supercells were truncated with the "cell_slab truncation" and "cell_box truncation" methods. We employed a 20 Ry energy cutoff for the epsilon matrix (epsilon cutoff), as well as a 20 Ry screened Coulomb cutoff. We used 5000 bands for 15x15x15 $Å^3$ cells and the equivalent number in other cells (scaled according to volume). For √3x√3 R30° cells we use a 3x3x1 Monkhorst-Pack k-mesh, while for 3x3 cells we use a 2x2x1 k-mesh. The $\mathbf{q} \rightarrow 0$ limit for metallic slabs was treated using a wavefunction with twice as dense a k-mesh but fewer unoccupied states. The static remainder method was used to help with convergence of the self-energy [11].

We performed one-shot GW calculations using a projection approach. To do this, we obtain the DFT converged wavefunction for both the slab ( $|\Psi_{slab}\rangle$ ) and the isolated molecule in the same geometry as in the slab ( $|\varphi_{mol}\rangle$ ). The slab wavefunction is used to construct the dielectric matrix ε and the self-energy operator Σ(E) as in normal GW calculations, but in the sigma output step, we utilize the WFN_outer facility that comes with the BerkeleyGW package [9], to evaluate the self-energy expectation values within the molecular wavefunction basis as in Eq. 1. More details are given in Section 2.

The image charge model used here is similar to Ref. [12]. The image plane position is taken to be 0.9 Å above the Au(111) surface [13]. The molecular orbital charge distribution was obtained using SIESTA [14] calculations of gas-phase molecules with the same geometry as their corresponding hybrid slabs, and mulliken populations at each atom are used to compute the image charge energy. For thiolate/Au systems, the molecular orbital distribution was obtained by passivating the thiolate group with the cleaved hydrogen atom. For molecules with metal contacts on both sides, a Mathematica worksheet was used to take into account infinite number of images. Visualization of all geometries in this work was achieved by the XCrysDen software [15].

**2) Details of molecular GW self-energy projection calculation**

In order to evaluate the self energy in the basis of the orbitals of the isolated molecule, the eqp_outer_corrections flag must be used in the sigma input file in BerkeleyGW.

When this flag is set, the sigma executable will look for WFN_outer, the wavefunctions used for evaluating the self-energy correction. WFN_outer will be the wavefunctions of the isolated molecule in this case.

Also, the sigma executable will look for eqp_outer.dat, that contains the DFT starting points for the eigenvalues, as well as the corrected energies at which the self energy is to be evaluated. We run sigma twice. In the first run, eqp_outer.dat contains the DFT levels of the molecular orbital peaks in the slab calculation (this is described below). The corrected energies are also set to be the same as these DFT starting levels. The sigma run will then generate a new eqp_outer.dat which replaces the corrected energies with the computed quasiparticle energies. Sigma is then run a second time, reading this new eqp_outer.dat. In this run, we should see in the output that eqp1 (the final quasiparticle value) is very close to ecor (the energy value read in from eqp_outer.dat, at which the sigma operator is evaluated).

The above will ensure that we are evaluating sigma using the molecular wavefunctions, and that the energies at which we are to evaluate sigma are not those of the isolated molecule, but are those of the molecular peaks in the slab calculation. We also need to make sure that the mean-field exchange-correlation term is computed using the molecular wavefunctions and not the slab wavefunctions. To do so, we do not copy vxc.dat (the matrix elements of the exchange-correlation potential) into the sigma folder but copy the exchange-correlation potential file VXC instead. This will make the sigma executable compute the required matrix elements from the exchange-correlation potential file VXC with the new WFN_outer wavefunctions basis.

For the above to work, the BerkeleyGW code ( Version 1.0.6 ) needs a few minor modifications. Firstly, BerkeleyGW will automatically check that the WFN_outer headings are the same as the WFN_inner headings. This check needs to be removed. Secondly, a new routine must be added to read the DFT starting levels (elda in the sigma output) from eqp_outer.dat, as explained above.

To get the initial DFT eigenvalue starting levels for the eqp_outer.dat file, we utilize the wavefunction projection utility wfn_dotproduct.x that comes with BerkeleyGW. The projection is done at Γ-point only, which outputs the squared modulus overlap weights of i-th molecular level upon all the hybrid slab wavefunction basis manifold and sums up to 1:

$$w_J^i = \left|\langle \varphi^i_{mol} | \Psi^J_{slab} \rangle\right|^2, \quad \sum_{J=1}^{N_{slab}} w_J^i = 1. \quad \text{(S1)}$$

By plotting the weights distribution of one particular MO level across the energy axis, we can readily identify an appropriate DFT energy point for this molecular state taken at the highest weighting factor point. In some cases, the MO can have weights scattered across a large energy range without any particular prominent peaks, which we take as spectroscopically undefined. The comparison of these DFT starting points with DFT molecular projected density of states (PDOS) plot reveals that they corresponding accordingly to respective molecular peaks in DFT PDOS.

### 3) GW convergence of Au(111) work function

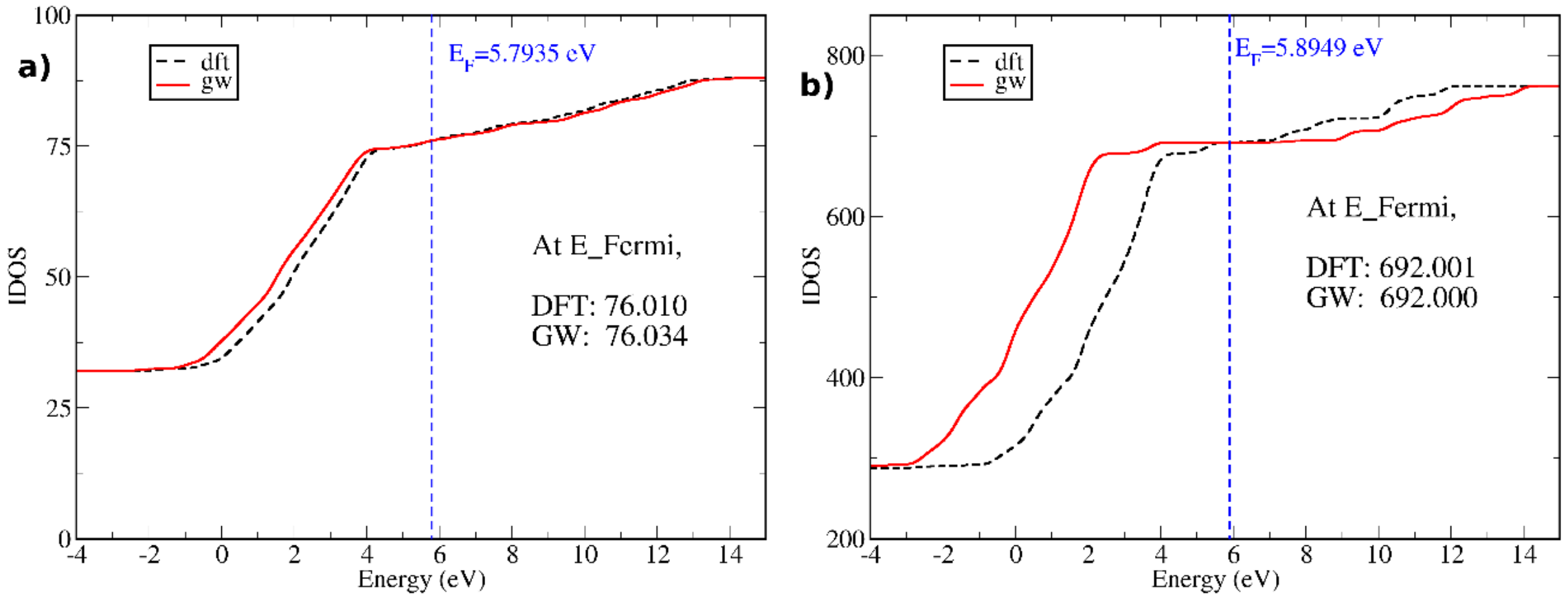

**Figure S1.** (a) Integrated density of states (IDOS) of Au(111) from GW and DFT PBE. DFT PBE predicts an accurate work function for Au(111) (5.3 eV). So if the IDOS from GW and PBE match well at the DFT Fermi level, GW will also give the same Fermi level, hence the same metal work function. Calculation details: 1x1 cell, 4 layer Au(111) slab cell with the $19e^-$ PBE norm-conserving Au pseudopotential optimized lattice constant at 8.06 Bohr, 13 Å vacuum space, 20 Ry epsilon cutoff and screened Coulomb cutoff, 60 Ry bare Coulomb cutoff, K-mesh/q-mesh 6x6x1, fine mesh 12x12x1 for approaching the $\mathbf{q} \to 0$ limit, slab Coulomb truncation. (b) Similar to a) but using the Wigner-Seitz box Coulomb truncation method on the 3x3 Au(111) cell with 4 layers of gold, calculated at the Γ point only. It is seen that this truncation does not affect the precision of our GW work function prediction.

**Table S1:** Comparison of results from different treatment of the $\mathbf{q} \to 0$ limit for bare Au(111) system as well as in Geo. **3**. For the Au(111) cell, we compare the difference between GW Fermi level and DFT Fermi level, i.e. $\mathrm{E_F^{GW}} - \mathrm{E_F^{DFT}}$; for Geo. **3**, we compare the difference between GW HOMO level and DFT Fermi level, i.e. $\mathrm{E_{HOMO}^{GW}} - \mathrm{E_F^{DFT}}$. We see that the change in the GW HOMO level for different treatments of the $\mathbf{q} \to 0$ limit was completely due to a corresponding GW Fermi level shift in an equivalent treatment of the $\mathbf{q} \to 0$ limit for the Au(111) slab. Different fine meshes are used to approach the $\mathbf{q} \to 0$ limit. With no Coulomb truncation, an averaging procedure is done in a region of the Brillouin Zone near $\mathbf{q} = 0$. It is important to note that the reference Fermi level of the GW spectrum has to be carefully calibrated against the DFT Fermi level. Otherwise, a rigid shift of the whole spectrum might result in incorrect level alignment predictions.

| Au(111) 1x1 cell, 4 layer slab | $\mathrm{E_F^{GW}} - \mathrm{E_F^{DFT}}$ (eV) | | Geo. **3**, 3x3 Au(111) cell with BDA | $\mathrm{E_{HOMO}^{GW}} - \mathrm{E_F^{DFT}}$ (eV) |
|---|---|---|---|---|
| 6x6x1 q-mesh, 12x12x1 fine-mesh, w/ slab truncation | **0** | | 2x2x1 q-mesh, 4x4x1 fine-mesh, w/ slab truncation | **-1.30** |
| 6x6x1 q-mesh, 18x18x1 fine-mesh, w/ slab truncation | **-0.37** | | 2x2x1 q-mesh, 6x6x1 fine-mesh, w/ slab truncation | **-1.66** |
| 6x6x1 q-mesh, 12x12x1 fine-mesh, no Coulomb truncation | **-0.65** | | 2x2x1 q-mesh, 4x4x1 fine-mesh, no Coulomb truncation | **-1.98** |

## 4) GW convergence of gas-phase molecular orbital levels

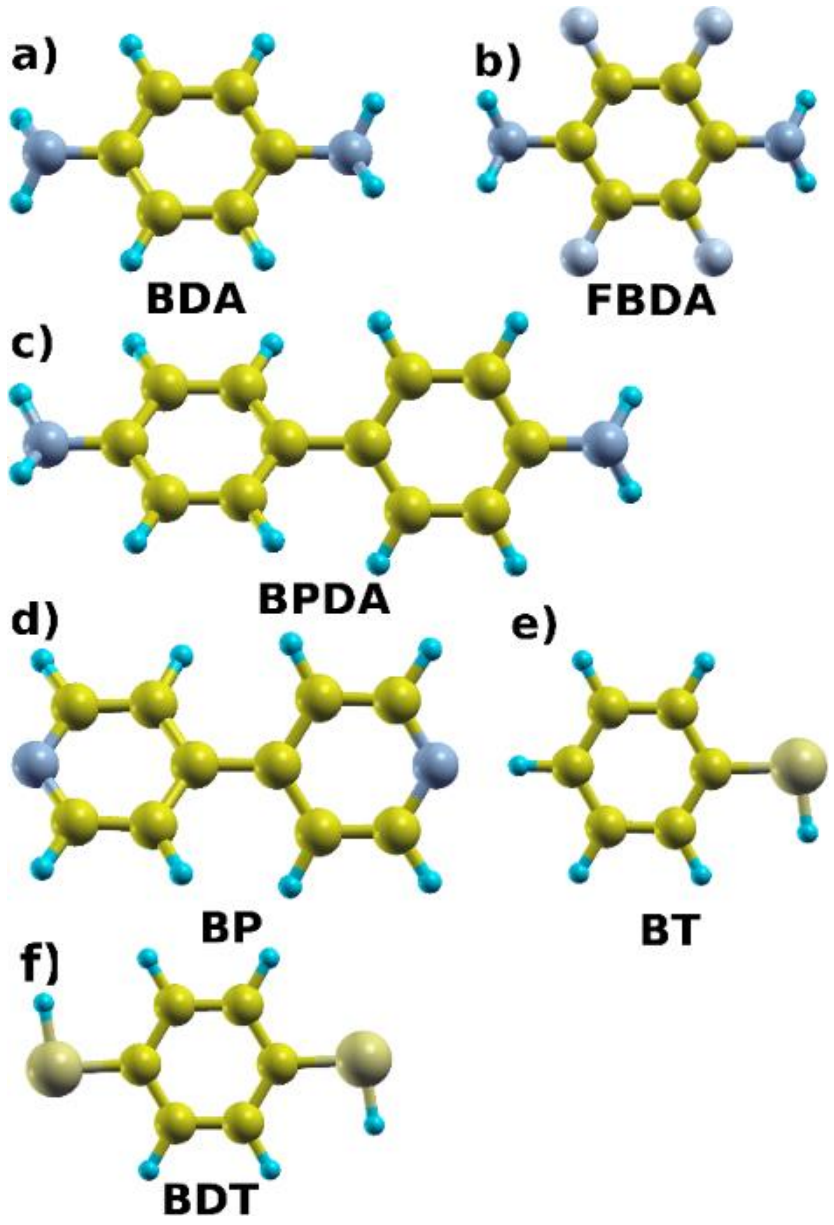

**Figure S2:** Molecules investigated in this work. BDA: BenzeneDiAmine; FBDA: Fluorinated BenzeneDiAmine; BPDA: BiPhenylDiAmine; BP: BiPyridine; BT: BenzeneThiol; BDT: BenzeneDiThiol.

**Table S2:** Converged GW results for the studied gas-phase single molecules. ΔE's refer to GW correction upon their DFT corresponding levels. $E_{gap}$ = IP – EA. Energy units are in eV. We note that the numbers reported here are obtained using a 24 Ry epsilon cutoff, 5000 bands for 15x15x15 Å$^3$ cells (following previous literature [16]), and a box Coulomb truncation scheme [9]. However, similar to Ref. [16], we obtain essentially the same results with a 20 Ry epsilon cutoff, which is used in our slab calculations.

| | IP($G_0W_0$) | EA($G_0W_0$) | ΔE(HOMO) | ΔE(LUMO) | $E_{gap}$($G_0W_0$) |
|---|---|---|---|---|---|
| BDA | 7.16 | -1.01 | -2.91 | +2.04 | 8.17 |
| FBDA | 7.85 | -0.37 | -2.97 | +1.90 | 8.22 |
| BPDA | 6.95 | -0.48 | -2.53 | +1.65 | 7.43 |
| BP | 9.32 | +0.99 | -3.27 | +1.89 | 8.33 |
| BT | 8.73 | -0.50 | -3.15 | +2.08 | 9.22 |
| BDT | 7.91 | -0.38 | -2.64 | +2.28 | 8.29 |

## 5) Off-diagonal elements of GW self-energy evaluated in the BDA MO basis for Geometry 4

**Table S3:** Off-diagonal elements of GW self-energy evaluated in the BDA MO basis for Geometry **4**. Only MOs close to the frontier orbitals are considered, as orbitals further away have even smaller probability of mixing with the frontier MOs. Results are shown for both the slab and box truncation schemes. The self-energy for each entry is evaluated at the energy corresponding to the MO identified by the row index of the entry.

| Slab GW | HOMO-1 | HOMO | LUMO | LUMO+1 |
|---|---|---|---|---|
| HOMO-1 | -13.8720 | -0.0052 -i*0.0021 | 0.0003+i*0.0074 | -0.0003 +i*0.0002 |
| HOMO | -0.0052 +i*0.0020 | -15.3778 | -0.0013-i*0.0548 | 0.0514 -i*0.0243 |
| LUMO | -0.0002-i*0.0112 | -0.0005 +i*0.0552 | -10.2048 | -0.1436 +i*0.0675 |
| LUMO+1 | -0.0003-i*0.0002 | 0.0481+i*0.0228 | -0.1443-i*0.0678 | -4.3570 |
| | | | | |
| Box GW | HOMO-1 | HOMO | LUMO | LUMO+1 |
| HOMO-1 | -15.9543 | 0.0002-i*0.0009 | -0.0014 +i*0.0136 | 0.0000-i*0.0000 |
| HOMO | 0.0002 +i*0.0009 | -17.4108 | -0.0019 -i*0.0000 | 0.0924-i*0.0432 |
| LUMO | -0.0027-i*0.0226 | -0.0018 +i*0.0000 | -10.2060 | -0.1349 +i*0.0628 |
| LUMO+1 | 0.0001 -i*0.0002 | 0.0888 +i*0.0416 | -0.1356 -i*0.0632 | -4.6174 |

**6) Off-diagonal elements of GW self-energy evaluated in the BP MO basis for Geometry 7**

**Table S4:** Off-diagonal elements of GW self-energy evaluated in the BP MO basis for Geometry **7**. Note that HOMO level at the interface is not well-defined here. Other details follow those in Table S3.

| Slab GW | HOMO-1 | LUMO | LUMO+1 |
|---|---|---|---|
| HOMO-1 | -17.7891 | -0.0034-i*0.0095 | 0.0919-i*0.0428 |
| LUMO | -0.0032+i*0.0095 | -13.7422 | -0.0005-i*0.0041 |
| LUMO+1 | 0.0922+i*0.0430 | -0.0005+i*0.0041 | -10.6064 |
| | | | |
| Box GW | HOMO-1 | LUMO | LUMO+1 |
| HOMO-1 | -20.0532 | -0.0007+i*0.0001 | 0.0874-i*0.0409 |
| LUMO | -0.0007-i*0.0001 | -16.0462 | -0.0001+i*0.0001 |
| LUMO+1 | 0.0876+i*0.0411 | -0.0001-i*0.0000 | -10.3741 |

## 7) Off-diagonal elements of GW self-energy evaluated in the BT MO basis for Geometry 11

**Table S5:** Off-diagonal elements of GW self-energy evaluated in the BT MO basis for Geometry **11**. Note that HOMO level at the interface is not well-defined here. Other details follow those in Table S3.

| Slab GW | HOMO-1 | HOMO | LUMO | LUMO+1 |
|---|---|---|---|---|
| HOMO-1 | -13.8048 | -0.0060 -i*0.0048 | 0.0190 -i*0.0057 | 0.0048 +i*0.0125 |
| HOMO | -0.0060 +i*0.0049 | -13.7622 | 0.0241 -i*0.0441 | 0.5737 +i*0.3368 |
| LUMO | 0.0172 +i*0.0051 | 0.0241 +i*0.0441 | -10.4989 | 0.0018 -i*0.0679 |
| LUMO+1 | 0.0053 -i*0.0137 | 0.5831 -i*0.3423 | 0.0018 +i*0.0678 | -11.1179 |
| | | | | |
| Box GW | HOMO-1 | HOMO | LUMO | LUMO+1 |
| HOMO-1 | -15.8129 | -0.0063 -i*0.0050 | 0.0220 -i*0.0103 | 0.0050 +i*0.0141 |
| HOMO | -0.0063 +i*0.0051 | -15.8501 | 0.0240 -i*0.0434 | 0.6618 +i*0.3916 |
| LUMO | 0.0084 +i*0.0051 | 0.0239 +i*0.0432 | -10.5077 | 0.0022 -i*0.0805 |
| LUMO+1 | 0.0061 -i*0.0170 | 0.6696 -i*0.3963 | 0.0022 +i*0.0805 | -11.2050 |

## 8) Other geometries considered for BDA molecular layer on Au(111)

The geometry used in Reference [6] was a face-on configuration in a 4x4 Au(111) cell, relaxed using DFT PBE without including van der Waals interactions. We have performed a geometry optimization of the same system using the vdW-DF2 functional and found that this coverage is likely to be too low (Figure S3). Geometry optimization using the vdW-DF2 functional found that the benzene ring is tilted 18.2° from the surface. Similar geometry optimization for the 3x3 Au(111) cell found a corresponding tilt angle of 24°, which is closer to the experimentally determined angle of 24°±10°. We also see that the amine group further from the surface is tilted toward the surface, away from the plane of the phenyl ring, by about 8°. This suggests that lower coverages would result in smaller tilt angles further from the experimental value, which is defined for "monolayer coverage" [6].

The geometry used in Reference [7] was a linear chain motif of BDA molecules on Au(111), which was experimentally observed at low temperature [17]. Using this same structure, we found a GW HOMO level of -0.75 eV and a DFT+Σ HOMO level of -2.58 eV (including the 0.3 eV intra-layer polarization as in Ref. [7]). The fact that GW predicts a HOMO level much closer to $E_F$ than the UPS value indicates that the linear chain geometry, which is stabilized by hydrogen bonds at 5 K [17], is not an appropriate geometry for the room temperature UPS experiment [6].

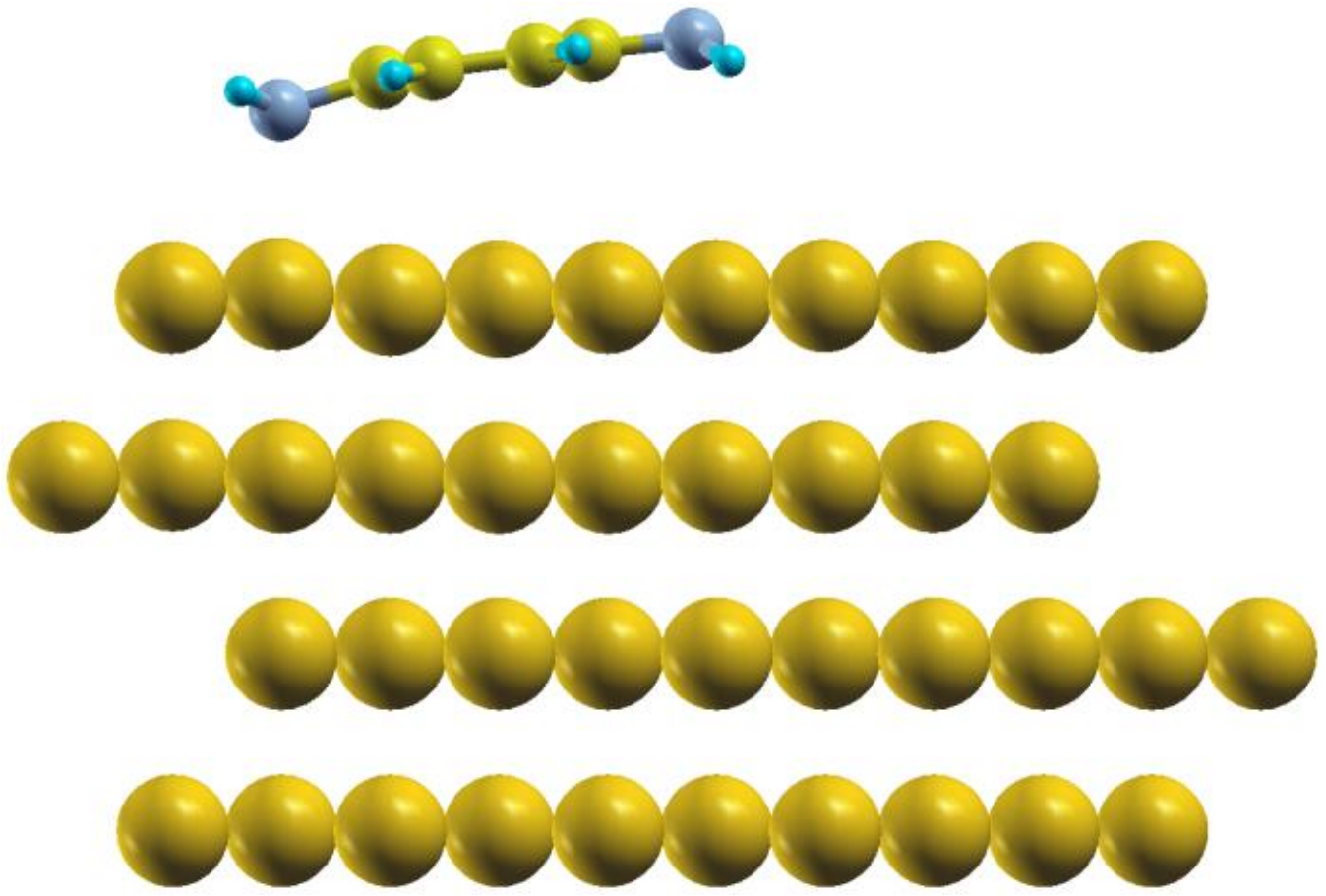

**Figure S3.** Optimized geometry of BDA in a 4x4 Au(111) cell in the face-on geometry (VASP, vdW-DF2 functional).

## 9) Geometries of metal/molecule/metal junction structures

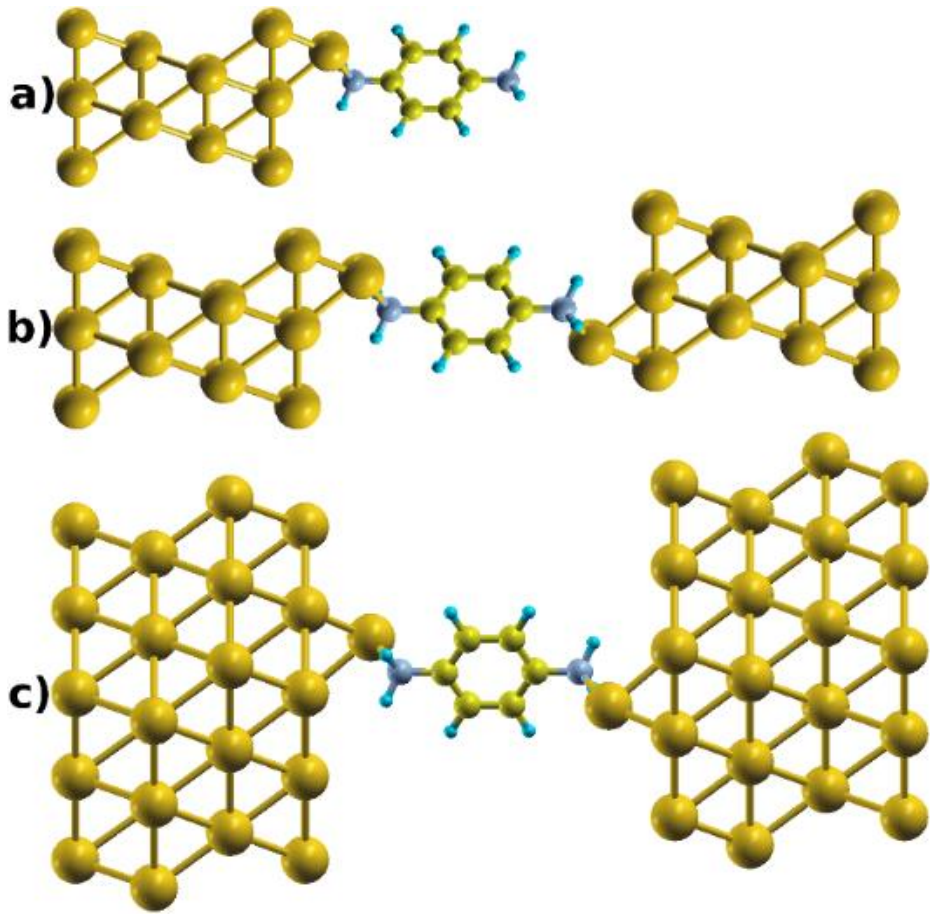

**Figure S4.** a) BDA vertical up-right geometry in √3x√3 R30° Au(111) cell; b) Au(111) /BDA/Au(111) junction structure with √3x√3 R30° cell; c) Au(111)/BDA/Au(111) junction structure with 3x3cell.

## 10) Decomposed self-energy contributions of MO levels from GW versus static-COHSEX results: Comprehensive result

**Table S6:** Components of the GW and static COHSEX self-energy corrections for different geometries, together with the bare exchange term (X). SX is screened exchange, CH the Coulomb hole term. The GW self-energy is equal to difference between (SX+CH) and the mean-field exchange-correlation term. The X term is given for comparison. Units are in eV. The red, blue and green colors are used as a guide to the eye for the discussion in the main text.

| | Method | X | SX | CH |
|---|---|---|---|---|
| BDA HOMO in Geo. **2** | Slab GW | -16.096 | -7.474 | -7.910 |
| | Slab static-COHSEX | -16.096 | -7.384 | -9.171 |
| | Box GW | -17.736 | -9.267 | -7.957 |
| | Box static-COHSEX | -17.736 | -9.266 | -9.689 |
| BDA HOMO in Geo. **4** | Slab GW | -15.904 | -7.286 | -8.033 |
| | Slab static-COHSEX | -15.904 | -7.162 | -9.356 |
| | Box GW | -17.474 | -9.158 | -7.923 |
| | Box static-COHSEX | -17.474 | -9.092 | -9.680 |
| BDA LUMO in Geo. **4** | Slab GW | -7.434 | -3.270 | -7.170 |
| | Slab static-COHSEX | -7.434 | -3.471 | -7.802 |
| | Box GW | -7.444 | -3.078 | -7.356 |
| | Box static-COHSEX | -7.444 | -3.304 | -8.115 |
| F4BDA HOMO in Geo. **5** | Slab GW | -16.753 | -8.094 | -8.068 |
| | Slab static-COHSEX | -16.753 | -8.048 | -9.355 |
| | Box GW | -18.440 | -9.897 | -8.058 |
| | Box static-COHSEX | -18.440 | -9.857 | -9.793 |
| BPDA HOMO in Geo. **6** | Slab GW | -15.054 | -7.323 | -7.600 |
| | Slab static-COHSEX | -15.054 | -7.376 | -8.806 |
| | Box GW | -16.754 | -9.027 | -7.662 |
| | Box static-COHSEX | -16.754 | -9.041 | -9.373 |
| BP HOMO-1 in Geo. **7** | Slab GW | -20.019 | -9.647 | -7.976 |
| | Slab static-COHSEX | -20.019 | -9.543 | -9.488 |
| | Box GW | -21.603 | -11.624 | -7.878 |
| | Box static-COHSEX | -21.603 | -11.597 | -9.846 |
| BP LUMO in Geo. **7** | Slab GW | -13.810 | -5.388 | -8.388 |
| | Slab static-COHSEX | -13.810 | -5.200 | -9.548 |
| | Box GW | -9.677 | -3.990 | -8.500 |
| | Box static-COHSEX | -9.677 | -4.221 | -9.438 |
| BT HOMO in Geo. **8** | Slab GW | -13.877 | -5.365 | -8.159 |
| | Slab static-COHSEX | -13.877 | -4.960 | -9.566 |
| | Box GW | -15.105 | -7.214 | -7.813 |
| | Box static-COHSEX | -15.105 | -6.898 | -9.564 |
| BT HOMO in Geo. **9** | Slab GW | -13.652 | -4.475 | -8.536 |
| | Slab static-COHSEX | -13.652 | -4.089 | -9.921 |
| | Box GW | -15.451 | -7.621 | -7.594 |
| | Box static-COHSEX | -15.451 | -7.534 | -9.301 |
| BT LUMO in Geo. **9** | Slab GW | -7.422 | -3.177 | -7.730 |
| | Slab static-COHSEX | -7.422 | -3.408 | -8.518 |
| | Box GW | -7.448 | -3.061 | -7.538 |
| | Box static-COHSEX | -7.448 | -3.323 | -8.297 |

## 11) Charge rearrangement and binding energy upon molecular adsorption on the Au(111) surface

**Table S7:** Number of electrons lost by the molecule (mol.) or the anchoring atom (N for amine and pyridine, S for thiol), and binding energy upon adsorption of the molecule on the Au(111) surface. The desorbed hydrogen atom from thiol is taken to carry a charge of 1 electron, and is taken into consideration in the final result below. (Geo.: Geometry) There is little net charge transfer from the molecule to the metal, but for thiols, there is significant local charge rearrangement in the molecule. The binding energy is computed as the difference in energy between the hybrid system and the sum of the energies of the gas phase molecule/radical and the Au surface.

| | Geo.**1** | Geo.**2** | Geo.**3** | Geo.**4** | Geo.**5** | Geo.**6** | Geo.**7** | Geo.**8** | Geo.**9** | Geo.**10** |
|---|---|---|---|---|---|---|---|---|---|---|
| Mol. | 0.086 | 0.182 | 0.159 | 0.193 | 0.142 | 0.177 | 0.048 | -0.117 | -0.167 | -0.132 |
| Anchor | -0.019 | -0.449 | -0.385 | 0.144 | 0.106 | 0.117 | 0.158 | -1.307 | -1.411 | -1.457 |
| Binding energy (eV) | -0.772 | -1.502 | -1.615 | -1.083 | -0.902 | -1.023 | -1.099 | -2.217 | -1.036 | -0.999 |

| | Geo.**4** @6L Au | Geo.**4** @4x4 Au | Geo.**9** @6L Au | Geo.**10** @6L Au | Geo. **11** |
|---|---|---|---|---|---|
| Mol. | 0.197 | 0.227 | -0.152 | -0.107 | -0.078 |
| Anchor | 0.129 | 0.159 | -1.484 | -1.442 | -1.334 |
| Binding energy (eV) | -1.091 | -1.123 | - | - | -2.189 |

## 12) Molecule/metal slab geometric parameters

**Table S8:** Summary of geometric parameters for Geometries **1** to **11**

| | Au-N(S) bond length (Å) | Au-N-C (Au-S-C) bond angle (°) |
|---|---|---|
| Geo. **1** | 2.609 | 136.86 |
| Geo. **2** | 2.359 | 119.87 |
| Geo. **3** | 2.448 | 112.5 |
| Geo. **4** | 2.338 | 114.0 |
| Geo. **5** | 2.338 | 114.0 |
| Geo. **6** | 2.338 | 114.2 |
| Geo. **7** | 2.154 | 120.0 |
| Geo. **8** | 2.279 | 105.74 |
| Geo. **9** | 2.270 | 179.90 |
| Geo. **10** | 2.276 | 179.82 |
| Geo. **11** | 2.489 | 132.058 |

Atomic geometric coordinates of BDA/Au systems from Geo. **1** to Geo. **4** are also listed below. Coordinates for Geo. **5 ~ 11** are listed at the end of this document after the Reference section.

XYZ format coordinates for Geo. **1**:

28

BDA/Au111 geo. 1

| | | | |
|---|---|---|---|
| C | 3.140518 | 1.457933 | 10.954336 |
| C | 4.460963 | 1.453559 | 13.472880 |
| C | 4.142933 | 2.650504 | 12.819166 |
| C | 3.523252 | 0.261119 | 11.567874 |
| C | 3.525667 | 2.652449 | 11.570825 |
| C | 4.140521 | 0.258851 | 12.816215 |
| H | 4.389917 | 3.604348 | 13.289313 |
| H | 3.274155 | -0.689438 | 11.092153 |
| H | 3.278212 | 3.604650 | 11.097572 |
| H | 4.385838 | -0.696546 | 13.284024 |
| H | 1.793448 | 2.298496 | 9.648112 |
| H | 5.549743 | 0.610949 | 14.995625 |
| H | 1.786475 | 0.626477 | 9.649273 |
| H | 5.551583 | 2.289944 | 14.997831 |
| N | 2.363610 | 1.460196 | 9.778419 |
| N | 5.026047 | 1.451323 | 14.757731 |
| Au | 0.240047 | -0.004960 | 7.177793 |
| Au | 2.802509 | 1.455866 | 7.206997 |
| Au | 5.283109 | 2.917041 | 7.177696 |
| Au | 1.808581 | 0.009923 | 4.762676 |
| Au | 1.808580 | 2.901915 | 4.762701 |
| Au | 4.312146 | 1.455923 | 4.762151 |
| Au | 0.840531 | 1.455842 | 2.366993 |
| Au | 3.362124 | 0.000000 | 2.366993 |
| Au | 3.362124 | 2.911685 | 2.366993 |
| Au | 0.000000 | 0.000000 | 0.000000 |
| Au | 2.521593 | 1.455842 | 0.000000 |
| Au | 5.043187 | 2.911685 | 0.000000 |

XYZ format coordinates for Geo. **2**:

==================================================

52

BDA/Au111 Geo. 2

| | | | |
|---|---|---|---|
| C | 4.761413451 | 4.019274501 | 10.522261754 |
| C | 2.620895606 | 3.443966919 | 12.262450780 |
| C | 3.946629270 | 3.451639331 | 12.723128482 |
| C | 3.451380398 | 3.970882823 | 10.059268094 |
| C | 5.001791441 | 3.737030845 | 11.864924817 |
| C | 2.393848463 | 3.694654121 | 10.902427675 |
| H | 4.148516089 | 3.250588215 | 13.771324166 |
| H | 3.257517136 | 4.149955505 | 9.004022563 |
| H | 6.017168085 | 3.762421231 | 12.252915793 |
| H | 1.388867942 | 3.658540452 | 10.489816377 |
| H | 6.733942585 | 4.147050554 | 9.981433813 |
| H | 0.692925395 | 2.941910612 | 12.718771470 |
| H | 5.803247145 | 5.361520897 | 9.365644656 |
| H | 1.772841755 | 2.788861356 | 14.005180670 |
| N | 5.812986221 | 4.366152204 | 9.611924720 |
| N | 1.561194696 | 3.246078170 | 13.131982881 |
| Au | 0.964117351 | -0.348343041 | 7.742372139 |
| Au | 4.530236087 | 1.030773970 | 7.687033493 |
| Au | 7.141760993 | 0.161546092 | 7.333429351 |
| Au | 2.061175236 | 2.157113304 | 7.524095839 |
| Au | 5.767629850 | 3.550809856 | 7.398783007 |
| Au | 8.365987446 | 2.611175034 | 7.352368745 |
| Au | 3.766769803 | 5.469249307 | 7.103858434 |
| Au | 7.871989956 | 6.345623619 | 7.378606960 |
| Au | 10.100949264 | 4.767738419 | 7.051396140 |
| Au | 0.628253963 | 1.975131362 | 4.857349387 |
| Au | 4.359172220 | 1.190983249 | 5.010798781 |
| Au | 6.740363878 | 2.519484037 | 4.977290905 |
| Au | 2.860036358 | 3.546187156 | 5.119587743 |
| Au | 5.298448399 | 4.889206840 | 4.896292938 |
| Au | 8.182069279 | 4.993620986 | 4.899221627 |
| Au | 3.838905079 | 7.338305521 | 4.851147983 |
| Au | 6.682169373 | 7.428123119 | 4.979732177 |
| Au | 10.318582343 | 6.681125101 | 5.036800014 |
| Au | 1.507965000 | 0.870624000 | 2.462496000 |
| Au | 4.523894000 | 0.870624000 | 2.462496000 |
| Au | 7.539824000 | 0.870624000 | 2.462496000 |
| Au | 3.015929000 | 3.482495000 | 2.462496000 |
| Au | 6.031859000 | 3.482495000 | 2.462496000 |
| Au | 9.047789000 | 3.482495000 | 2.462496000 |
| Au | 4.523894000 | 6.094367000 | 2.462496000 |
| Au | 7.539824000 | 6.094367000 | 2.462496000 |
| Au | 10.555753000 | 6.094367000 | 2.462496000 |
| Au | 0.000000000 | 0.000000000 | 0.000000000 |
| Au | 3.015929000 | 0.000000000 | 0.000000000 |
| Au | 6.031859000 | 0.000000000 | 0.000000000 |

Au 1.507965000 2.611871000 0.000000000

Au 4.523894000 2.611871000 0.000000000

Au 7.539824000 2.611871000 0.000000000

Au 3.015929000 5.223743000 0.000000000

Au 6.031859000 5.223743000 0.000000000

Au 9.047789000 5.223743000 0.000000000

XYZ format coordinates for Geo. **3**:

=================================================

```
52
BDA/Au111 Geo. 3
 C      2.730355     2.522484    10.070164
 C      5.422029     2.522759    10.925045
 C      4.735766     3.730650    10.704206
 C      3.410923     1.314465    10.276190
 C      3.409572     3.730663    10.279437
 C      4.737030     1.314724    10.700964
 H      5.252808     4.683311    10.855477
 H      2.903116     0.363397    10.082341
 H      2.900752     4.681685    10.088120
 H      5.255090     0.362164    10.849473
 H      0.874735     3.363748     9.711786
 H      7.264235     1.673163    11.253534
 H      0.875384     1.680238     9.712041
 H      7.263835     3.372660    11.255507
 N      1.420943     2.522159     9.519558
 N      6.728310     2.522613    11.407515
Au     -0.011328    -0.021809     7.093648
Au      1.432353     2.519782     7.071760
Au      2.916514     5.057871     7.063904
Au      2.914489    -0.022022     7.066985
Au      4.382766     2.519448     7.050173
Au      5.824010     5.041463     7.104085
Au      5.822985    -0.003273     7.101907
Au      7.253042     2.521951     7.093094
Au      8.720242     5.058489     7.096251
Au     -1.460520    -0.835989     4.700900
Au     -0.003236     1.679730     4.724558
Au      1.449835     4.198133     4.724686
```

```
Au    1.450752   -0.844558    4.710308
Au    2.901150    1.682475    4.725055
Au    4.369957    4.207025    4.706791
Au    4.371940   -0.840565    4.703816
Au    5.826000    1.672797    4.702075
Au    7.277906    4.202270    4.714493
Au   10.190898    5.883718    2.366993
Au    7.279213    0.840531    2.366993
Au    8.735055    3.362125    2.366993
Au    4.367527    5.883718    2.366993
Au    1.455842    0.840531    2.366993
Au    2.911685    3.362125    2.366993
Au    7.279212    5.883718    2.366993
Au    4.367527    0.840531    2.366993
Au    5.823370    3.362125    2.366993
Au    0.000000    0.000000    0.000000
Au    1.455843    2.521594    0.000000
Au    2.911685    5.043188    0.000000
Au    2.911685    0.000000    0.000000
Au    4.367527    2.521594    0.000000
Au    5.823370    5.043188    0.000000
Au    5.823370    0.000000    0.000000
Au    7.279213    2.521594    0.000000
Au    8.735055    5.043188    0.000000
```

XYZ format coordinates for Geo. **4**:

===========================================================

```
53
BDA/Au111 Geo.4
C        1.891566814    4.576038109   11.573503922
C        1.832116814    4.543150409   14.408967922
C        2.442723814    3.513916254   13.685161622
C        1.274174814    5.601394909   12.297214522
C        2.470823814    3.528774977   12.297576922
C        1.245677014    5.585894609   13.684983222
H        2.905559814    2.687522649   14.219049422
H        0.803113014    6.424243409   11.765379422
H        2.947296814    2.707463029   11.768434522
H        0.763359114    6.403440209   14.215219522
H        2.598312814    4.012539629    9.744261222
H        1.123661014    5.089952909   16.242661122
H        1.782871814    5.467323309    9.739604922
```

| | | | |
|---|---|---|---|
| H | 1.956700814 | 3.648590329 | 16.235336122 |
| N | 1.866445814 | 4.560087409 | 10.173820022 |
| N | 1.866410814 | 4.560116309 | 15.809484122 |
| Au | 0.000000604 | 3.482494306 | 9.268659967 |
| Au | 0.000000000 | 0.000000000 | 7.201817163 |
| Au | -1.507964117 | 2.611870469 | 7.201817163 |
| Au | -3.015928234 | 5.223740938 | 7.201817163 |
| Au | 3.015928239 | 0.000000000 | 7.201817163 |
| Au | 1.507964122 | 2.611870469 | 7.201817163 |
| Au | 0.000000005 | 5.223740938 | 7.201817163 |
| Au | 6.031856478 | 0.000000000 | 7.201817163 |
| Au | 4.523892361 | 2.611870469 | 7.201817163 |
| Au | 3.015928244 | 5.223740938 | 7.201817163 |
| Au | 0.000000000 | 1.741247675 | 4.842043812 |
| Au | -1.507964117 | 4.353118144 | 4.842043812 |
| Au | -3.015928234 | 6.964988613 | 4.842043812 |
| Au | 3.015928239 | 1.741247675 | 4.842043812 |
| Au | 1.507964122 | 4.353118144 | 4.842043812 |
| Au | 0.000000005 | 6.964988613 | 4.842043812 |
| Au | 6.031856478 | 1.741247675 | 4.842043812 |
| Au | 4.523892361 | 4.353118144 | 4.842043812 |
| Au | 3.015928244 | 6.964988613 | 4.842043812 |
| Au | 1.507964721 | 0.870623837 | 2.462496078 |
| Au | 0.000000604 | 3.482494306 | 2.462496078 |
| Au | -1.507963513 | 6.094364775 | 2.462496078 |
| Au | 4.523892960 | 0.870623837 | 2.462496078 |
| Au | 3.015928843 | 3.482494306 | 2.462496078 |
| Au | 1.507964726 | 6.094364775 | 2.462496078 |
| Au | 7.539821199 | 0.870623837 | 2.462496078 |
| Au | 6.031857082 | 3.482494306 | 2.462496078 |
| Au | 4.523892965 | 6.094364775 | 2.462496078 |
| Au | 0.000000000 | 0.000000000 | 0.000000000 |
| Au | -1.507964117 | 2.611870469 | 0.000000000 |
| Au | -3.015928234 | 5.223740938 | 0.000000000 |
| Au | 3.015928239 | 0.000000000 | 0.000000000 |
| Au | 1.507964122 | 2.611870469 | 0.000000000 |
| Au | 0.000000005 | 5.223740938 | 0.000000000 |
| Au | 6.031856478 | 0.000000000 | 0.000000000 |
| Au | 4.523892361 | 2.611870469 | 0.000000000 |
| Au | 3.015928244 | 5.223740938 | 0.000000000 |

## 13) DFT and GW bandstructure for the BDA layer as in Geo. 4

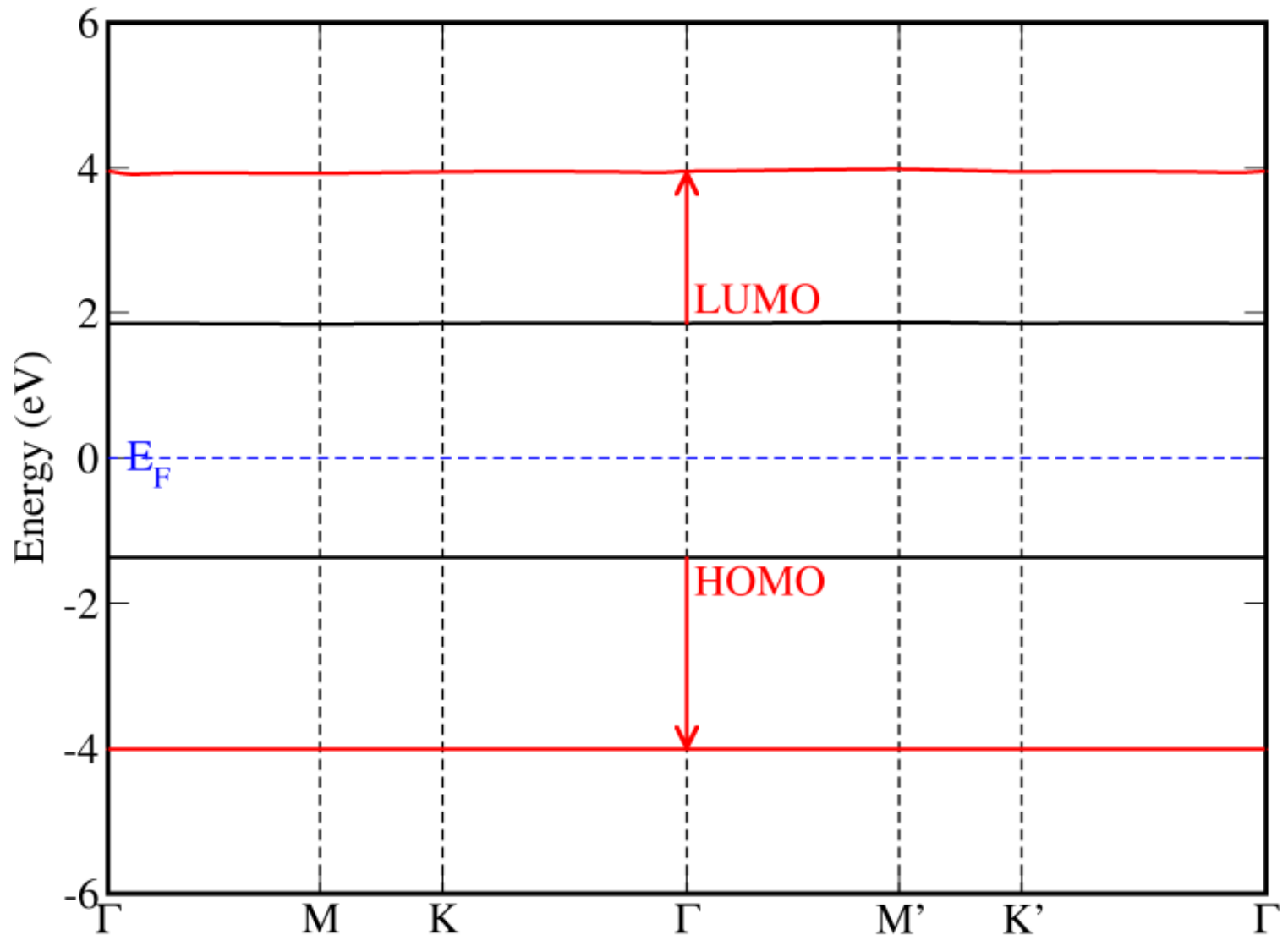

**Figure S5.** DFT (Black) and GW (Red) bandstructure for BDA molecular layer as in Geo. **4**, along the k-path Γ → M → K → Γ → M′ → K′ → Γ calculated with 4x4 k-mesh sampling, and 12 Ry epsilon cutoff in GW calculation. Bandwidth for HOMO and LUMO bands are 0.00055 eV (0.00066 eV) and 0.025 eV (0.073 eV) respectively, for the DFT (GW) results. They are not dispersive across the Brillouin zone. GW HOMO-LUMO gap at Γ for the layer is 7.959 eV, while the corresponding Wigner-Seitz truncation gives 8.169 eV. So the intra-molecular layer polarization effect is only 0.21 eV.

## 14) GW projected molecular level dispersion across the Brillouin zone

We also performed slab truncation GW projection calculations of the HOMO and LUMO levels of BDA as in Geo. 4 at another three k-points in the Brillouin zone (BZ). The results are summarized in the table below (with respect to the Fermi level). There is seen to be little dispersion even for the projected levels from the molecule-metal hybrid slab. This indicates the molecular band gap renormalization we observe is not from BZ dispersion of the adsorbed molecular layer. A molecular layer GW projected bandstructure is currently beyond the reach of our computational capabilities.

**Table S9** Slab GW HOMO and LUMO levels for Geometry **4**

| (eV) | Γ | K=(0.0, 0.5, 0.0) | K=(0.5, 0.0, 0.0) | K=(0.5, 0.5, 0.0) |
|---|---|---|---|---|
| HOMO | -1.344 | -1.345 | -1.345 | -1.346 |
| LUMO | +4.230 | +4.208 | +4.207 | +4.251 |

## 15) Calculated image charge correction energy for the hybridized slabs

**Table S10:** Image charge model correction energy for various MO levels in the hybridized geometries investigated in this work. Here, we compare the image charge correction we have computed with that assuming a unit of point charge (point charge method) at the ring center or in the middle of the molecule (Geometries 6 and 7 where there are two rings). The image plane is taken to be 0.9 Å above the Au(111) surface.

| | Geo.**1** **HOMO** | Geo.**2** **HOMO** | Geo.**3** **HOMO** | Geo. **4** **HOMO** | Geo.**5** **HOMO** | Geo.**6** **HOMO** |
|---|---|---|---|---|---|---|
| Image charge correction used in the main text (eV) | 0.9612 | 1.2181 | 1.3059 | 0.8174 | 0.8129 | 0.5913 |
| Distance of point charge from image plane (Å) | 4.126 | 3.094 | 2.515 | 4.889 | 4.892 | 7.043 |
| Point charge method (eV) | 0.8725 | 1.1630 | 1.4314 | 0.7363 | 0.7358 | 0.5112 |
| | Geo.**7** **LUMO** | Geo.**8** **HOMO** | Geo.**9** **HOMO** | Geo.**10** **HOMO** | Geo.**11** **HOMO** | |
| Image charge correction used in the main text (eV) | 0.5631 | 0.8639 | 0.7729 | 0.5937 | 2.3382 | |
| Distance of point charge from image plane (Å) | 6.864 | 5.447 | 6.590 | 6.576 | 3.957 | |
| Point charge method (eV) | 0.5239 | 0.6609 | 0.5463 | 0.5474 | 0.9097 | |

## 16) Projected density of states together with our DFT starting points marked for Geo.'s 1-11 investigated in this work.

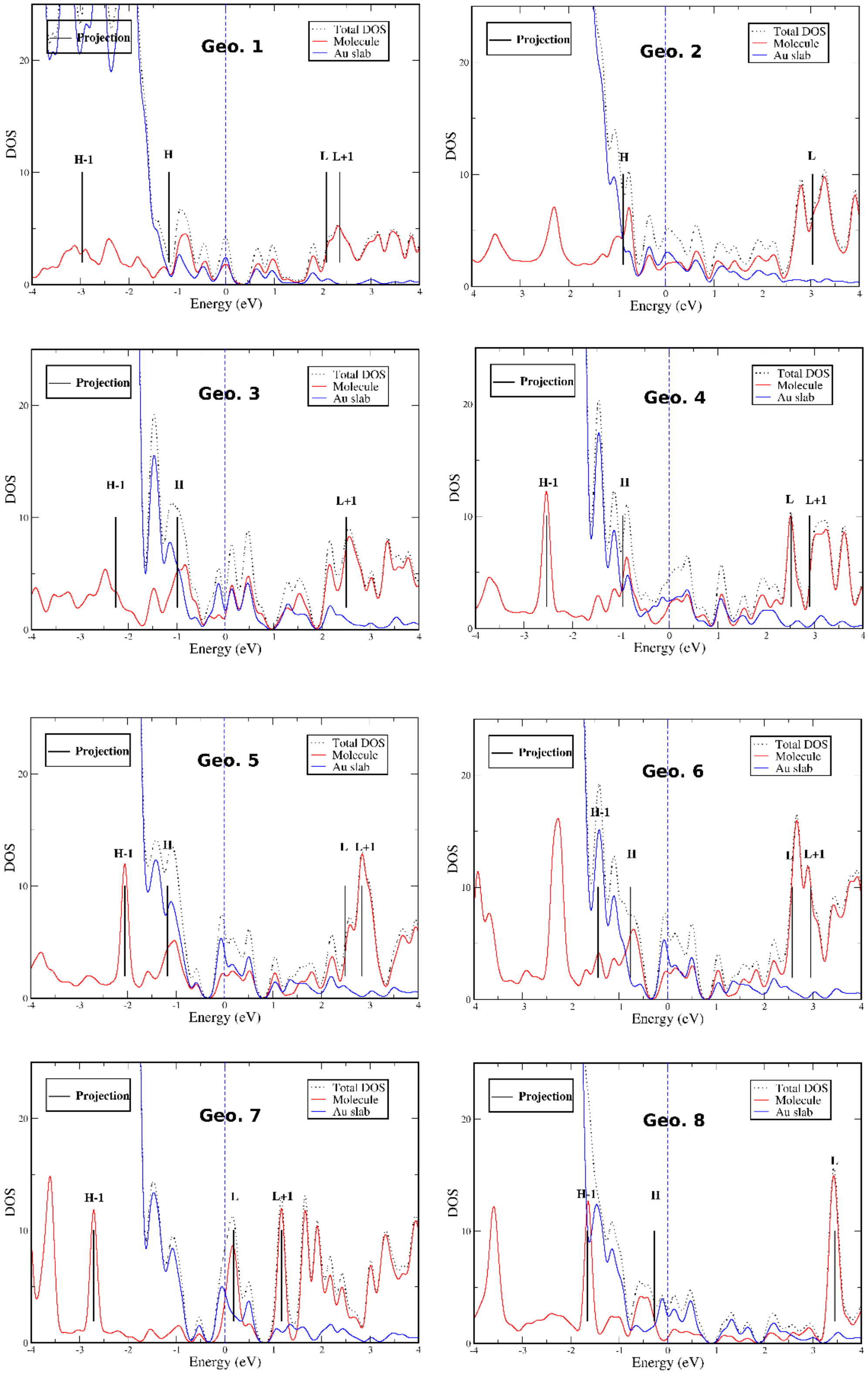

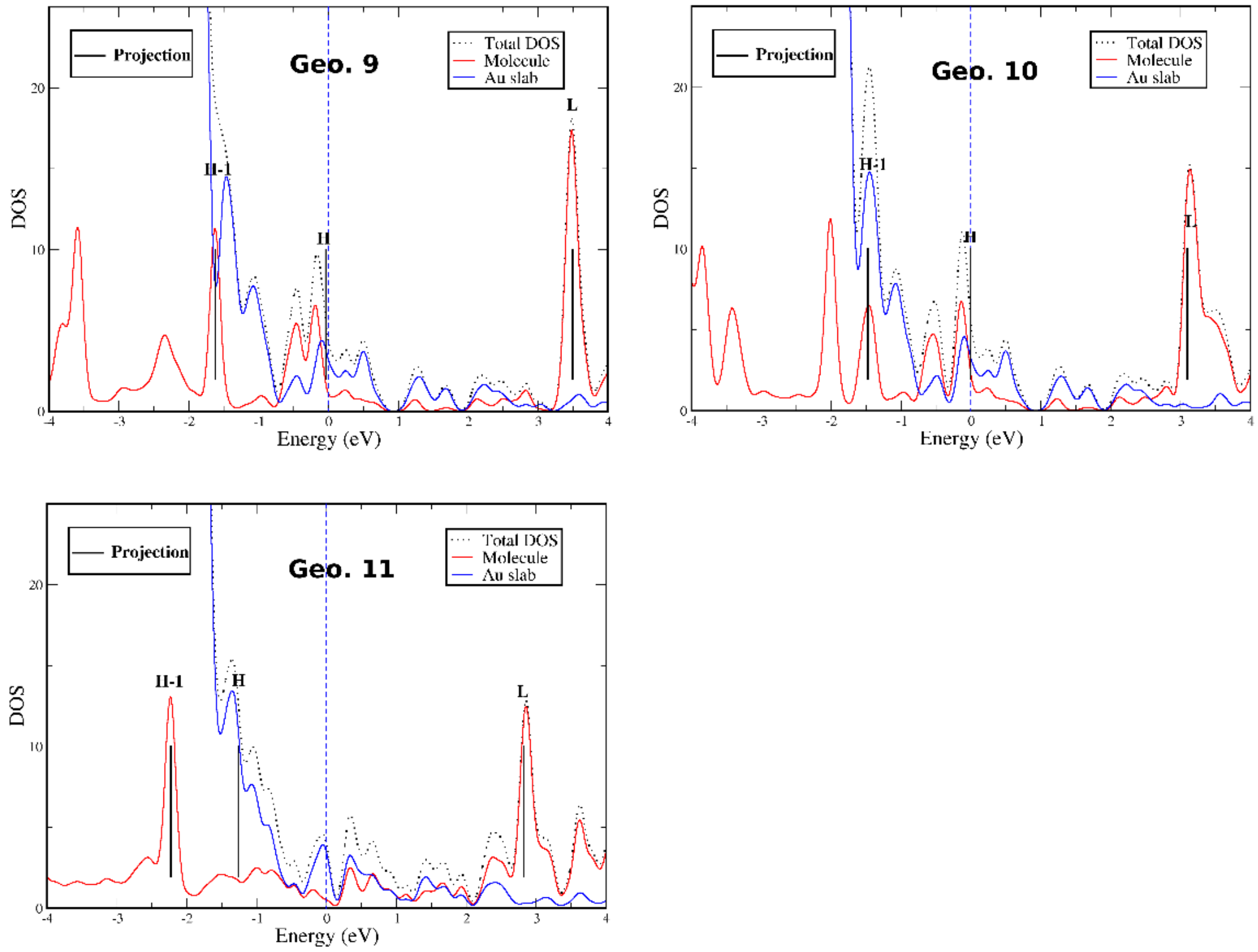

**Figure S6.** Projected density of states plots, with our wavefunction based projected DFT HOMO-1 (**H-1**), HOMO (**H**), LUMO (**L**), and LUMO+1 (**L+1**) starting point levels marked on the same plot.

## 17) Table S11 is given in the next page.

Spectroscopically undefined levels (without a well-defined peak) are omitted.

1 17) Summary of energy levels of geometries investigated in this work.

2 Table S11: Energy levels for various structures. Fermi levels are absolute values for each calculation, while all the rest data are given with respective to the molecule/metal slab Fermi level of each case.

| 3 | System | DFT HOMO | GW HOMO (Slab) | GW HOMO (Box) | ImageMeth HOMO | DFT LUMO | GW LUMO (Slab) | GW LUMO (Box) | ImageMeth LUMO | E_Fermi |
|---|---|---|---|---|---|---|---|---|---|---|
| 4 | BDA-Tl/3x3Au111 NoAd PBE-D2 (2) | -0.873 | -1.3392 | -3.181 | -2.5698 | 3.0364 | 3.7004 | 3.5394 | 3.8748 | 3.8352 |
| 5 | BDA-Tl/3x3Au111 W/Ad PBE-D2 | -0.2852 | -0.7426 | -2.6012 | -2.221 | 2.9602 | 3.35 | 3.2152 | 4.0993 | 4.0491 |
| 6 | BDA-Tl/Sqr(3)^2 R30 Au NoAd VdW (1) | -1.1726 | -1.6428 | -4.2112 | -3.1254 | 2.0744 | 2.7894 | None | 3.2501 | 5.2874 |
| 7 | BDA-Tl/Sqr(3)^2 R30 Au W/Ad VdW | -2.5634 | -3.4123 | -5.5623 | -4.8104 | 0.619 | Wrg Occ. | None | 2.0305 | 6.6385 |
| 8 | BDA-Face/3x3Au111 NoAd VdW (3) | -0.99 | -1.2988 | None | -2.5981 | N.A. | N.A. | None | None | 5.9661 |
| 9 | BDA-Face/3x3Au111 W/Ad QE | -1.653 | -2.2559 | None | -3.8025 | 1.8643 | 2.6849 | None | 3.1377 | 4.3414 |
| 10 | BDA-Face/3x3Au111 NoAd QE | -0.8712 | -1.1779 | None | -2.5155 | N.A. | N.A. | None | None | 4.7646 |
| 11 | BDA-Vert/3x3Au111 W/Ad QE (4) | -0.9582 | -1.3443 | -3.1074 | -3.0548 | 2.5157 | 4.2303 | 4.236 | 3.8222 | 3.6168 |
| 12 | Ibid (Stat.-COHSEX) | -0.9582 | -2.5428 | -4.7966 | as above | 2.5157 | 3.3967 | 3.2504 | as above | ibid |
| 13 | BDA-Vert/3x3Au111 W/Ad VdW | -1.059 | -1.4883 | -3.3575 | n.a. | 2.8243 | 2.8919 | 2.5205 | n.a. | 4.2737 |
| 14 | BDA-LinChain/Au111 NoAd @G. Li | -0.571 | -0.7501 | None | N.A. | 2.9374 | 4.2039 | None | N.A. | 4.2096 |
| 15 | F4BDA-Vert/3x3Au111 w/Ad VdW (5) | -1.1804 | -1.7128 | -3.5074 | -3.3345 | 2.4934 | 3.5888 | 3.3388 | 3.662 | 4.3361 |
| 16 | BPDA-Vert/3x3Au111 w/Ad VdW (6) | -0.7649 | -1.1672 | -2.9356 | -2.7026 | 2.5765 | 3.6686 | 3.338 | 3.6684 | 3.522 |
| 17 | Bpyd-Vert/3x3Au111 w/Ad VdW (7) | N.A. | N.A. | N.A. | N.A. | 0.1782 | 0.3992 | 1.6856 | 1.5081 | 4.0253 |
| 18 | Ibid (Stat.-COHSEX) | N.A. | N.A. | N.A. | as above | 0.1782 | Wrg Occ | 0.5158 | as above | ibid |
| 19 | BDA-Vert/4x4Au111 4L W/Ad VdW (4'') | -0.5856 | N.A. | -2.3749 | -2.6822 | 2.8016 | N.A. | 4.2129 | 4.1081 | 3.0178 |
| 20 | BDA-Vert/3x3Au111 6L W/Ad VdW (4') | -0.8052 | N.A. | -2.7759 | -2.9018 | 2.6282 | N.A. | 4.2318 | 3.9347 | 5.4439 |
| 21 | BDA-Vert/Sqr(3)^2 R30 W/ Ad VdW | -1.1381 | -1.7853 | -4.4266 | -3.2358 | 1.5964 | Wrg Occ. | Wrg Occ. | 2.9032 | 6.0106 |
| 22 | Au/BDA-Vt/Au Sqr(3)^2 R30 Ad Junc | -1.2142 | -1.484 | -4.0579 | -2.9532 | 1.2977 | Wrg Occ. | Wrg Occ. | 2.2947 | 9.6787 |
| 23 | Au/BDA-Vt/Au 3x3 Ad Junc | -0.5855 | None | -2.146 | -2.3212 | 2.6448 | None | 4.0031 | 3.641 | 6.5264 |
| 24 | Au/BP-Vt/Au 3x3 Ad Junc | N.A. | N.A. | N.A. | N.A. | 0.3879 | None | 0.8386 | 1.4885 | 6.5612 |
| 25 | BDA/Sqr(3)^2 R30 NoAd PBE-D2 (1') | -1.087 | -1.6984 | None | -3 | 2.3667 | 2.9918 | None | 3.4684 | 4.2709 |
| 26 | BT/Sqr(3)^2 R30 W/Ad PBE-D2 | -0.3716 | -1.0261 | None | -2.2327 | 2.6472 | 3.6602 | None | 3.956 | 4.5505 |
| 27 | BT-Tl/3x3Au111 4L w/Ad VdW (8) | -0.273 | -0.7212 | -2.2252 | -2.5551 | 3.4541 | 4.8332 | 4.8044 | 4.8734 | 3.4517 |
| 28 | Ibid (Stat.-COHSEX) | -0.273 | -1.7232 | -3.6594 | as above | 3.4541 | 3.8739 | 3.7652 | as above | ibid |
| 29 | BT-Rlx/3x3Au111 4L NoAd VdW (11) | -1.2639 | -1.8048 | -3.5968 | -2.0717 | 2.8252 | 4.4558 | 4.4492 | 3.9909 | 3.6974 |
| 30 | Ibid (Stat.-COHSEX) | -1.2639 | -2.9642 | -5.2787 | as above | 2.8252 | 3.4946 | 3.4063 | as above | ibid |
| 31 | BT-Rlx/3x3Au111 4L w/Ad VdW (9) | -0.0362 | -0.2711 | -2.4786 | -2.4093 | 3.4958 | 4.8948 | 5.2038 | 5.0271 | 3.4169 |
| 32 | Ibid (Stat.-COHSEX) | -0.0362 | -1.2705 | -4.096 | as above | 3.4958 | 3.8763 | 4.1818 | as above | ibid |
| 33 | BT-Rlx/3x3Au111 6L Ad VdW | -0.1586 | None | -2.15 | -2.5317 | 3.4257 | None | 4.9231 | 4.957 | 5.2395 |
| 34 | BDT-Vert/3x3Au111 4L W/Ad VdW (10) | -0.0044 | -0.3444 | -2.5532 | -2.0557 | 3.1 | 4.3967 | 4.741 | 4.8325 | 3.6124 |
| 35 | BDT-Vert/3x3Au111 6L W/Ad VdW | -0.1127 | None | -2.2746 | -2.164 | 3.0327 | None | 4.4643 | 4.7652 | 5.414 |

XYZ format coordinates for Geo. **5**:

```
===============================================
53
FBDA/Au111 Geo.5
C        3.346936459    1.933185327   11.504639189
C        3.284275559    1.898938327   14.371124289
C        3.891758959    0.882371327   13.631277989
C        2.740061959    2.950213427   12.244251689
C        3.922006159    0.899107327   12.245656989
C        2.709002959    2.933199927   13.630192789
```

| | | | |
|---|---|---|---|
| F | 4.466391559 | -0.144357673 | 14.304729489 |
| F | 2.174253959 | 3.981608427 | 11.570697789 |
| F | 4.529808959 | -0.109429673 | 11.573615989 |
| F | 2.107283159 | 3.945006327 | 14.302213689 |
| H | 4.046689959 | 1.358908327 | 9.690505489 |
| H | 2.583575959 | 2.465780227 | 16.188327589 |
| H | 3.217785559 | 2.825453727 | 9.687817089 |
| H | 3.428218359 | 1.008035327 | 16.187054589 |
| N | 3.322287359 | 1.918124327 | 10.117536089 |
| N | 3.310993659 | 1.914196327 | 15.758096589 |
| Au | 1.455842 | 0.840531 | 9.212376 |
| Au | 0.000000 | 0.000000 | 7.145576 |
| Au | 1.455842 | 2.521594 | 7.145576 |
| Au | 2.911685 | 5.043187 | 7.145576 |
| Au | 2.911685 | 0.000000 | 7.145576 |
| Au | 4.367527 | 2.521594 | 7.145576 |
| Au | 5.823370 | 5.043187 | 7.145576 |
| Au | 5.823370 | 0.000000 | 7.145576 |
| Au | 7.279213 | 2.521594 | 7.145576 |
| Au | 8.735055 | 5.043187 | 7.145576 |
| Au | -1.455842 | -0.840530 | 4.740462 |
| Au | 0.000000 | 1.681062 | 4.740462 |
| Au | 1.455842 | 4.202656 | 4.740462 |
| Au | 1.455842 | -0.840530 | 4.740462 |
| Au | 2.911685 | 1.681062 | 4.740462 |
| Au | 4.367527 | 4.202656 | 4.740462 |
| Au | 4.367527 | -0.840530 | 4.740462 |
| Au | 5.823370 | 1.681062 | 4.740462 |
| Au | 7.279212 | 4.202656 | 4.740462 |
| Au | -2.911684 | -1.681061 | 2.366993 |
| Au | -1.455842 | 0.840531 | 2.366993 |
| Au | 0.000000 | 3.362125 | 2.366993 |
| Au | 0.000000 | -1.681061 | 2.366993 |
| Au | 1.455842 | 0.840531 | 2.366993 |
| Au | 2.911685 | 3.362125 | 2.366993 |
| Au | 2.911685 | -1.681061 | 2.366993 |
| Au | 4.367527 | 0.840531 | 2.366993 |
| Au | 5.823370 | 3.362125 | 2.366993 |
| Au | 0.000000 | 0.000000 | 0.000000 |
| Au | 1.455842 | 2.521594 | 0.000000 |
| Au | 2.911685 | 5.043187 | 0.000000 |
| Au | 2.911685 | 0.000000 | 0.000000 |
| Au | 4.367527 | 2.521594 | 0.000000 |
| Au | 5.823370 | 5.043187 | 0.000000 |
| Au | 5.823370 | 0.000000 | 0.000000 |

```
Au       7.279213    2.521594    0.000000
Au       8.735055    5.043187    0.000000
```

XYZ format coordinates for Geo. **6**:

```
=============================================

63
BPDA/Au111 Geo. 6
C        3.355147993   1.929198734  11.507437954
C        3.335166093   1.898626734  15.824901554
C        2.746321293   2.959346734  12.235939054
C        2.251890783   2.406469734  16.555333154
C        2.744683193   2.943040834  13.621835254
C        2.236940743   2.406404734  17.941023154
C        3.340460893   1.904830734  14.351695054
C        3.315899293   1.887271734  18.669003154
C        3.937884493   0.875307734  13.611186954
C        4.408231393   1.382604734  17.951214154
C        3.946577693   0.880544734  12.224902154
C        4.411557793   1.389212734  16.565165554
H        2.279821273   3.786867134  11.705902554
H        1.383887413   2.786858534  16.024913754
H        2.291950583   3.775274234  14.152350454
H        1.373523833   2.802698134  18.470975154
H        4.387785293   0.035310734  14.132204554
H        5.269483193   0.994900734  18.490933154
H        4.405103293   0.055064734  11.685146354
H        5.285158793   1.009420734  16.043815654
H        4.025957493   1.359131734   9.659574754
H        3.180130793   2.809911734   9.668415554
H        2.427767873   1.997757734  20.508727154
H        3.940838193   1.269188734  20.517403154
N        3.322287393   1.918023734  10.117536054
N        3.327639693   1.925412734  20.058623154
Au       1.455842    0.840531    9.212376
Au       0.000000    0.000000    7.145576
Au       1.455842    2.521594    7.145576
Au       2.911685    5.043187    7.145576
Au       2.911685    0.000000    7.145576
Au       4.367527    2.521594    7.145576
Au       5.823370    5.043187    7.145576
Au       5.823370    0.000000    7.145576
Au       7.279213    2.521594    7.145576
Au       8.735055    5.043187    7.145576
Au      -1.455842   -0.840530    4.740462
```

```
Au       0.000000    1.681062    4.740462
Au       1.455842    4.202656    4.740462
Au       1.455842   -0.840530    4.740462
Au       2.911685    1.681062    4.740462
Au       4.367527    4.202656    4.740462
Au       4.367527   -0.840530    4.740462
Au       5.823370    1.681062    4.740462
Au       7.279212    4.202656    4.740462
Au      -2.911684   -1.681061    2.366993
Au      -1.455842    0.840531    2.366993
Au       0.000000    3.362125    2.366993
Au       0.000000   -1.681061    2.366993
Au       1.455842    0.840531    2.366993
Au       2.911685    3.362125    2.366993
Au       2.911685   -1.681061    2.366993
Au       4.367527    0.840531    2.366993
Au       5.823370    3.362125    2.366993
Au       0.000000    0.000000    0.000000
Au       1.455842    2.521594    0.000000
Au       2.911685    5.043187    0.000000
Au       2.911685    0.000000    0.000000
Au       4.367527    2.521594    0.000000
Au       5.823370    5.043187    0.000000
Au       5.823370    0.000000    0.000000
Au       7.279213    2.521594    0.000000
Au       8.735055    5.043187    0.000000
```

XYZ format coordinates for Geo. **7**:

```
==========================================================

57
BP/Au111 Geo. 7
C          2.609926793    0.821302516   12.056242343
C          0.448298316    0.339810507   17.767854339
C          2.649631142    0.833870231   13.435745220
C          0.393356000    0.323579453   16.379715619
C          1.460313362    0.842881731   14.176526920
C          1.464719776    0.842743837   15.643683747
C          0.266784301    0.848458909   13.442073723
C          2.540800797    1.358348077   16.375483503
C          0.298820733    0.845534869   12.062005361
C          2.492748462    1.338967827   17.763943210
H          3.516483889    0.798050830   11.459771484
H         -0.371864948   -0.076305610   18.349660787
```

| | | | |
|---|---|---|---|
| H | 3.611368932 | 0.810875717 | 13.934398978 |
| H | -0.462652251 | -0.117662284 | 15.879364189 |
| H | -0.691999414 | 0.878410069 | 13.946053359 |
| H | 3.395731724 | 1.798424833 | 15.872675457 |
| H | -0.611111453 | 0.864182907 | 11.470558529 |
| H | 3.315778389 | 1.752749059 | 18.343104391 |
| N | 1.452606489 | 0.827735032 | 11.365970727 |
| N | 1.471978177 | 0.839046221 | 18.467563258 |
| Au | 1.455842000 | 0.840531000 | 9.212376000 |
| Au | 0.000000000 | 0.000000000 | 7.145576000 |
| Au | 1.455842000 | 2.521594000 | 7.145576000 |
| Au | 2.911685000 | 5.043187000 | 7.145576000 |
| Au | 2.911685000 | 0.000000000 | 7.145576000 |
| Au | 4.367527000 | 2.521594000 | 7.145576000 |
| Au | 5.823370000 | 5.043187000 | 7.145576000 |
| Au | 5.823370000 | 0.000000000 | 7.145576000 |
| Au | 7.279213000 | 2.521594000 | 7.145576000 |
| Au | 8.735055000 | 5.043187000 | 7.145576000 |
| Au | -1.455842000 | -0.840530000 | 4.740462000 |
| Au | -0.000000000 | 1.681062000 | 4.740462000 |
| Au | 1.455842000 | 4.202656000 | 4.740462000 |
| Au | 1.455842000 | -0.840530000 | 4.740462000 |
| Au | 2.911685000 | 1.681062000 | 4.740462000 |
| Au | 4.367527000 | 4.202656000 | 4.740462000 |
| Au | 4.367527000 | -0.840530000 | 4.740462000 |
| Au | 5.823370000 | 1.681062000 | 4.740462000 |
| Au | 7.279212000 | 4.202656000 | 4.740462000 |
| Au | -2.911684000 | -1.681061000 | 2.366993000 |
| Au | -1.455842000 | 0.840531000 | 2.366993000 |
| Au | -0.000000000 | 3.362125000 | 2.366993000 |
| Au | -0.000000000 | -1.681061000 | 2.366993000 |
| Au | 1.455842000 | 0.840531000 | 2.366993000 |
| Au | 2.911685000 | 3.362125000 | 2.366993000 |
| Au | 2.911685000 | -1.681061000 | 2.366993000 |
| Au | 4.367527000 | 0.840531000 | 2.366993000 |
| Au | 5.823370000 | 3.362125000 | 2.366993000 |
| Au | 0.000000000 | 0.000000000 | 0.000000000 |
| Au | 1.455842000 | 2.521594000 | 0.000000000 |
| Au | 2.911685000 | 5.043187000 | 0.000000000 |
| Au | 2.911685000 | 0.000000000 | 0.000000000 |
| Au | 4.367527000 | 2.521594000 | 0.000000000 |
| Au | 5.823370000 | 5.043187000 | 0.000000000 |
| Au | 5.823370000 | 0.000000000 | 0.000000000 |
| Au | 7.279213000 | 2.521594000 | 0.000000000 |
| Au | 8.735055000 | 5.043187000 | 0.000000000 |

XYZ format coordinates for Geo. **8**:

=====================================================

```
49
BT/Au111 Geo. 8
 C     2.150145    -1.230141    14.477068
 C     3.115240    -0.239798    14.271947
 C     1.018261    -1.278469    13.655580
 C     1.811997     0.660244    12.432239
 C     0.843390    -0.339825    12.640182
 C     2.953773     0.698985    13.253397
 H     2.279572    -1.963974    15.275338
 H     4.003986    -0.199515    14.906179
 H     0.260375    -2.050132    13.811606
 H    -0.046466    -0.365081    12.006510
 H     3.709088     1.469532    13.082018
 S     1.587966     1.944259    11.238962
Au     1.465806     0.896358     9.218989
Au    -0.039413    -0.033752     7.150378
Au     1.458566     2.592555     7.053531
Au     2.911708     5.051297     7.108723
Au     2.941963    -0.020757     7.118739
Au     4.381521     2.531892     7.098454
Au     5.824805     5.037709     7.118122
Au     5.822330     0.004812     7.106397
Au     7.270741     2.529382     7.098300
Au     8.731442     5.053135     7.117217
Au    -1.441147    -0.828900     4.739646
Au    -0.001690     1.669536     4.742193
Au     1.451877     4.211063     4.714260
Au     1.449829    -0.831777     4.749635
Au     2.916351     1.681528     4.720239
Au     4.367039     4.207519     4.722154
Au     4.357423    -0.834021     4.729552
Au     5.823437     1.678444     4.717889
Au     7.281885     4.209522     4.722643
Au    -2.911684    -1.681061     2.366993
Au    -1.455842     0.840531     2.366993
Au     0.000000     3.362125     2.366993
Au     0.000000    -1.681061     2.366993
Au     1.455842     0.840531     2.366993
Au     2.911685     3.362125     2.366993
Au     2.911685    -1.681061     2.366993
```

```
Au     4.367527     0.840531     2.366993
Au     5.823370     3.362125     2.366993
Au     0.000000     0.000000     0.000000
Au     1.455843     2.521594     0.000000
Au     2.911685     5.043187     0.000000
Au     2.911685     0.000000     0.000000
Au     4.367527     2.521594     0.000000
Au     5.823370     5.043187     0.000000
Au     5.823370     0.000000     0.000000
Au     7.279213     2.521594     0.000000
Au     8.731442     5.053135     0.000000
```

XYZ format coordinates for Geo. **9**:

```
=====================================================

49
BT/Au111 Geo. 9
 C     1.426690     0.839128    15.979084
 C     1.437812     0.838811    13.179546
 C     0.219173     0.838967    15.272281
 C     0.214734     0.838797    13.879280
 C     2.639688     0.838943    15.281778
 C     2.655425     0.838801    13.888760
 H    -0.730831     0.838986    15.812101
 H    -0.722712     0.839139    13.320462
 H     3.585295     0.838949    15.829356
 H     3.597467     0.839153    13.337756
 H     1.422346     0.839351    17.071012
 S     1.442356     0.838610    11.425965
Au     1.450868     0.841494     9.156417
Au    -0.052733    -0.034792     7.091561
Au     1.453200     2.582392     7.096040
Au     2.907229     5.041691     7.093721
Au     2.958261    -0.034232     7.093453
Au     4.375841     2.526106     7.080925
Au     5.820825     5.029961     7.083813
Au     5.820347    -0.002465     7.089806
Au     7.265927     2.526024     7.081017
Au     8.734023     5.042040     7.093887
Au     2.920018     6.727657     4.719918
Au     0.002202     1.677897     4.721198
Au     1.454337     4.189516     4.719575
Au     5.821989     6.724917     4.718755
Au     2.906571     1.677815     4.721424
```

```
Au     4.361479     4.203880     4.712465
Au     8.723205     6.727948     4.720487
Au     5.821971     1.673987     4.711652
Au     7.282404     4.203845     4.712494
Au    10.190896     5.883717     2.366993
Au     7.279211     0.840531     2.366993
Au     8.735053     3.362125     2.366993
Au     4.367527     5.883717     2.366993
Au     1.455842     0.840531     2.366993
Au     2.911685     3.362125     2.366993
Au     7.279212     5.883717     2.366993
Au     4.367527     0.840531     2.366993
Au     5.823370     3.362125     2.366993
Au     0.000000     0.000000     0.000000
Au     1.455843     2.521594     0.000000
Au     2.911685     5.043187     0.000000
Au     2.911685     0.000000     0.000000
Au     4.367527     2.521594     0.000000
Au     5.823370     5.043187     0.000000
Au     5.823370     0.000000     0.000000
Au     7.279213     2.521594     0.000000
Au     8.735055     5.043187     0.000000
```

XYZ format coordinates for Geo. **10**:

```
===========================================================

50
BDA/Au111 Geo. 10
 C     1.450287     0.839070    15.967575
 C     1.436239     0.839678    13.152452
 C     0.232534     0.839533    15.261061
 C     0.222492     0.839878    13.872808
 C     2.660110     0.839449    15.250630
 C     2.655280     0.839722    13.862775
 H    -0.714431     0.839709    15.807552
 H    -0.722960     0.840254    13.327235
 H     3.613864     0.839714    15.784193
 H     3.595005     0.840010    13.307438
 H     2.714658     0.835950    17.932905
 S     1.380408     0.838385    17.723131
 S     1.441391     0.837451    11.411168
Au     1.441553     0.837765     9.135625
Au    -0.057260    -0.039232     7.074550
Au     1.453108     2.584203     7.083807
Au     2.906261     5.039850     7.092668
```

| | | | |
|---|---|---|---|
| Au | 2.958892 | -0.036463 | 7.080940 |
| Au | 4.376632 | 2.525972 | 7.079947 |
| Au | 5.820942 | 5.029329 | 7.083890 |
| Au | 5.818905 | -0.002927 | 7.088006 |
| Au | 7.265678 | 2.525811 | 7.080161 |
| Au | 8.734281 | 5.041441 | 7.093029 |
| Au | 2.917288 | 6.725523 | 4.715117 |
| Au | 0.001077 | 1.678335 | 4.716452 |
| Au | 1.454314 | 4.190536 | 4.715515 |
| Au | 5.822172 | 6.722654 | 4.713393 |
| Au | 2.907832 | 1.678137 | 4.717274 |
| Au | 4.362059 | 4.203358 | 4.712223 |
| Au | 8.724665 | 6.726484 | 4.716837 |
| Au | 5.822145 | 1.674022 | 4.711135 |
| Au | 7.282074 | 4.203297 | 4.712359 |
| Au | 10.190896 | 5.883717 | 2.366993 |
| Au | 7.279211 | 0.840531 | 2.366993 |
| Au | 8.735053 | 3.362125 | 2.366993 |
| Au | 4.367527 | 5.883717 | 2.366993 |
| Au | 1.455842 | 0.840531 | 2.366993 |
| Au | 2.911685 | 3.362125 | 2.366993 |
| Au | 7.279212 | 5.883717 | 2.366993 |
| Au | 4.367527 | 0.840531 | 2.366993 |
| Au | 5.823370 | 3.362125 | 2.366993 |
| Au | 0.000000 | 0.000000 | 0.000000 |
| Au | 1.455843 | 2.521594 | 0.000000 |
| Au | 2.911685 | 5.043187 | 0.000000 |
| Au | 2.911685 | 0.000000 | 0.000000 |
| Au | 4.367527 | 2.521594 | 0.000000 |
| Au | 5.823370 | 5.043187 | 0.000000 |
| Au | 5.823370 | 0.000000 | 0.000000 |
| Au | 7.279213 | 2.521594 | 0.000000 |
| Au | 8.735055 | 5.043187 | 0.000000 |

XYZ format coordinates for Geo. **11**:

==============================================

48
BT/Au111 Geo.11

| | | | |
|---|---|---|---|
| C | 1.427023 | 0.835287 | 13.362285 |
| C | 1.444150 | 0.813753 | 10.575703 |
| C | 0.220572 | 0.827981 | 12.655337 |
| C | 0.219814 | 0.816530 | 11.259890 |
| C | 2.642057 | 0.828828 | 12.670319 |
| C | 2.660004 | 0.817531 | 11.274996 |
| H | -0.729934 | 0.830400 | 13.194242 |
| H | -0.718246 | 0.809103 | 10.698633 |
| H | 3.585868 | 0.831860 | 13.220914 |
| H | 3.604888 | 0.810943 | 10.725375 |
| H | 1.420272 | 0.844064 | 14.454275 |
| S | 1.456048 | 0.832010 | 8.793631 |
| Au | -0.144635 | -0.082733 | 7.132111 |
| Au | 1.456660 | 2.686360 | 7.116473 |
| Au | 2.916801 | 5.039582 | 7.135703 |
| Au | 3.058789 | -0.083550 | 7.130434 |
| Au | 4.397068 | 2.539729 | 7.073653 |
| Au | 5.824298 | 5.013067 | 7.073183 |
| Au | 5.824420 | 0.008593 | 7.137298 |
| Au | 7.251223 | 2.539947 | 7.073566 |
| Au | 8.732137 | 5.039474 | 7.135849 |
| Au | 2.918450 | 6.728941 | 4.730779 |
| Au | 0.015958 | 1.672855 | 4.765087 |
| Au | 1.456520 | 4.201386 | 4.723540 |
| Au | 5.824058 | 6.745527 | 4.768819 |
| Au | 2.897121 | 1.672792 | 4.765215 |
| Au | 4.358670 | 4.209696 | 4.703969 |
| Au | 8.730309 | 6.728533 | 4.729741 |
| Au | 5.824068 | 1.670957 | 4.703792 |
| Au | 7.289482 | 4.209714 | 4.704031 |
| Au | 10.190896 | 5.883717 | 2.366993 |
| Au | 7.279211 | 0.840531 | 2.366993 |
| Au | 8.735053 | 3.362125 | 2.366993 |
| Au | 4.367527 | 5.883717 | 2.366993 |
| Au | 1.455842 | 0.840531 | 2.366993 |
| Au | 2.911685 | 3.362125 | 2.366993 |
| Au | 7.279212 | 5.883717 | 2.366993 |
| Au | 4.367527 | 0.840531 | 2.366993 |
| Au | 5.823370 | 3.362125 | 2.366993 |
| Au | 0.000000 | 0.000000 | 0.000000 |
| Au | 1.455843 | 2.521594 | 0.000000 |
| Au | 2.911685 | 5.043187 | 0.000000 |
| Au | 2.911685 | 0.000000 | 0.000000 |
| Au | 4.367527 | 2.521594 | 0.000000 |
| Au | 5.823370 | 5.043187 | 0.000000 |
| Au | 5.823370 | 0.000000 | 0.000000 |
| Au | 7.279213 | 2.521594 | 0.000000 |
| Au | 8.735055 | 5.043187 | 0.000000 |